\documentclass[onecolumn,a4paper,11pt,oneside]{article} 
\usepackage[top=2.5cm,left=2.5cm,right=2.5cm,bottom=3cm]{geometry}    
\usepackage[nostamp]{draftwatermark} 
\SetWatermarkText{\bf DRAFT}
\SetWatermarkAngle{45}
\SetWatermarkScale{4}
\SetWatermarkLightness{0.85}     
\usepackage[parfill]{parskip}    
\usepackage{graphicx}
\usepackage{amssymb}
\usepackage{amsmath}
\usepackage{amsfonts}
\usepackage{amsthm}
\usepackage{epstopdf}
\usepackage[usenames,dvipsnames,svgnames,table]{xcolor}
\usepackage{tikz}
\usepackage{array}
\usepackage[T1]{fontenc}
\usepackage[utf8]{inputenc}
\usepackage{booktabs} 
\usepackage{array} 
\usepackage{paralist} 
\usepackage{listings} 
\usepackage{verbatim} 
\usepackage{subfig} 
\usepackage[colorlinks=true]{hyperref}
\hypersetup{colorlinks=true, linktocpage=true, linkcolor=Blue, citecolor=Green,
hypertexnames=true, urlcolor=MidnightBlue,pdftex}
\usepackage[english]{babel}
\usepackage{wrapfig}
\usepackage{multirow}
\usepackage{quoting}
\usepackage{rotating}
\quotingsetup{font=normalsize}
\theoremstyle{plain}

\usepackage{bm}
\usepackage{abstract}
\usepackage{natbib}
\usepackage{float}
\usepackage{textcomp} 	
\usepackage{tabularx} 
\usepackage{caption}
\usepackage{authblk} 
\usepackage{eso-pic} 
\usepackage{enumitem} 
\usepackage{lineno} 
\providecommand{\keywords}[1]{\textbf{{Key words: }} #1} 

\newcommand{\be}{\begin{equation}}
\newcommand{\ee}{\end{equation}}
\newcommand{\bsp}{\begin{split}}
\newcommand{\esp}{\end{split}}

\renewcommand{\Phi}{\varPhi}

\renewcommand{\Theta}{\varTheta}
\renewcommand{\Psi}{\varPsi}
\renewcommand{\Sigma}{\varSigma}

\renewcommand{\Delta}{\varDelta}
\renewcommand{\phi}{\varphi}
\renewcommand{\psi}{\varPsi}

\newcommand{\cha}{{\it charpente}}
\newcommand{\com}{{\it combles}}

\title{\LARGE{\bf Structural study of the\\Notre-Dame ancient  \textit{charpente}}\medskip\\\small{\textsc{Preprint}}}

\author{P. Vannucci\\\footnotesize{Université Paris-Saclay, UVSQ, CNRS, Laboratoire de Mathématiques de Versailles\\78035, Versailles, France.\\ \href{mailto:paolo.vannucci@uvsq.fr}{paolo.vannucci@uvsq.fr}}}

\begin{document}
\maketitle

\hrule
\begin{abstract}
The timber roofing structure ({\it charpente} or {\it combles}, in French) of the cathedral Notre-Dame, destroyed by the fire of April 15th, 2019, is studied. The aim is twofold: on the one hand, it is interesting to evaluate the structural behavior of the  original wooden structure in view of the reconstruction of the cathedral's roof. On the other hand, its structural analysis, never done before, can help to shed a light on the design process used by the masterbuilders of the XIIIth century, and to reconstruct, at least in part, the structural thought and knowledge of the ancient builders.

\end{abstract}
\medskip
\hrule
\bigskip


\section{Introduction}
\begin{figure}
\begin{center}
\includegraphics[width=.55\textwidth]{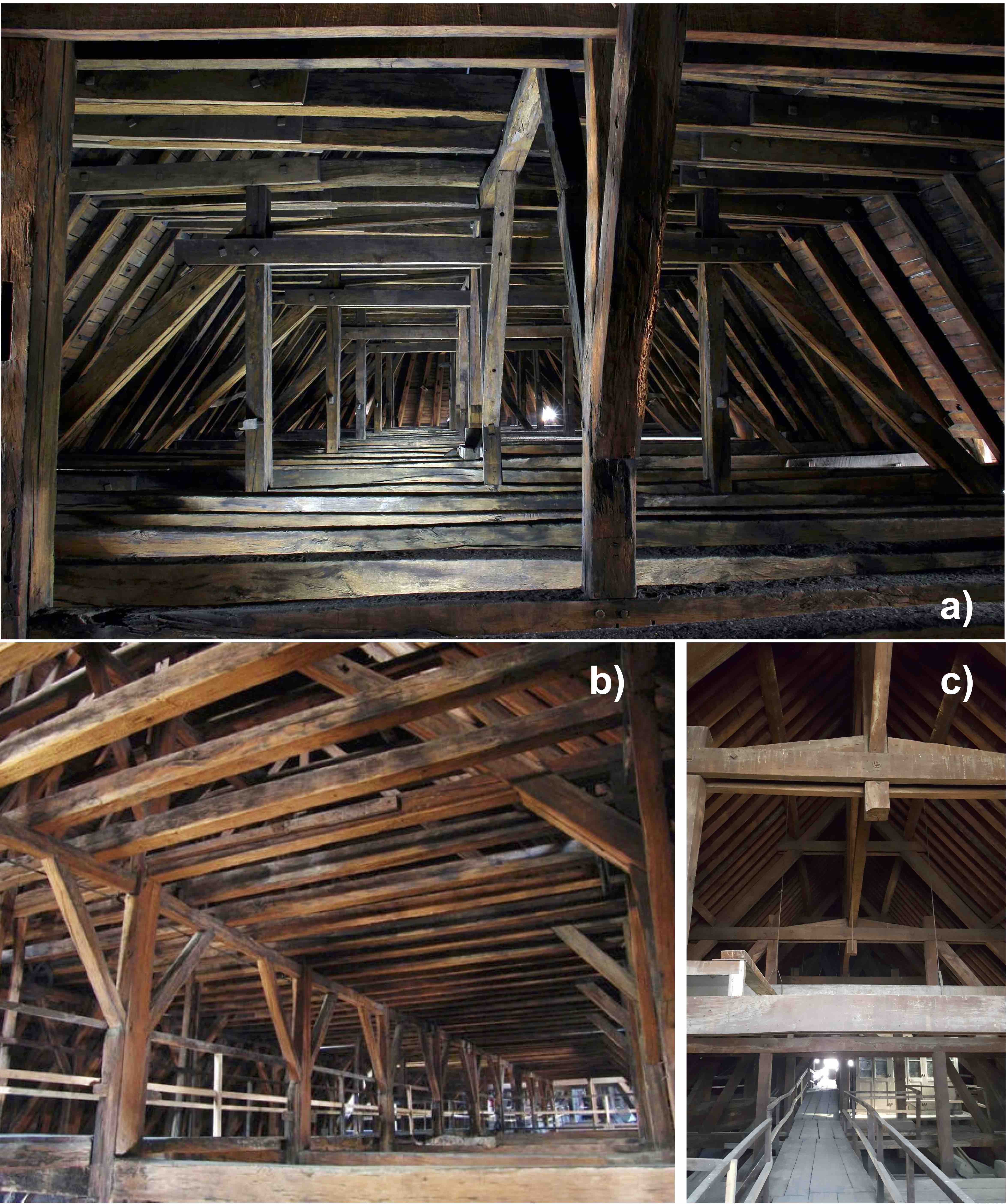}
\caption{Views of the  \com: a) in the choir (from \cite{fromont}, b) in the nave (from \cite{epaud1}, c) in the transept (by the author).}
\label{fig:1}
\end{center}
\end{figure}

\markboth{\textit{Preprint}}{\textit{Preprint}}
The fire occurred on April the 15th, 2019, at the Cathedral Notre-Dame of Paris  entirely destroyed its roofing structure. This was an impressive timber construction (in French, the \com\ or the {\it charpente}; these two terms will be used in the following as synonymous, though the first one more properly indicates the part of the building above the high vault of a cathedral, while the second one refers to any structure composed of ties, struts and beams), almost completely dating back to the XIIIth century, Fig. \ref{fig:1}. Actually, it was not constituted by a unique original structure, but at least by three distinct parts, built at different epochs, \cite{aubert2}, \cite{epaud2}:
\begin{enumerate}
\item the choir {\it charpente}, built after 1220, probably from 1225 to 1230;
\item the nave {\it charpente}, slightly subsequent, presumably built from 1230 to 1240;
\item the transept {\it charpente}, entirely rebuilt during the restoration campaign of  Lassus and Viollet-le-Duc after 1843, along with the spire and the first frames of the the nave and choir  nearby the spire.
\end{enumerate}

\begin{figure}
\begin{center}
\includegraphics[height=0.25\textheight]{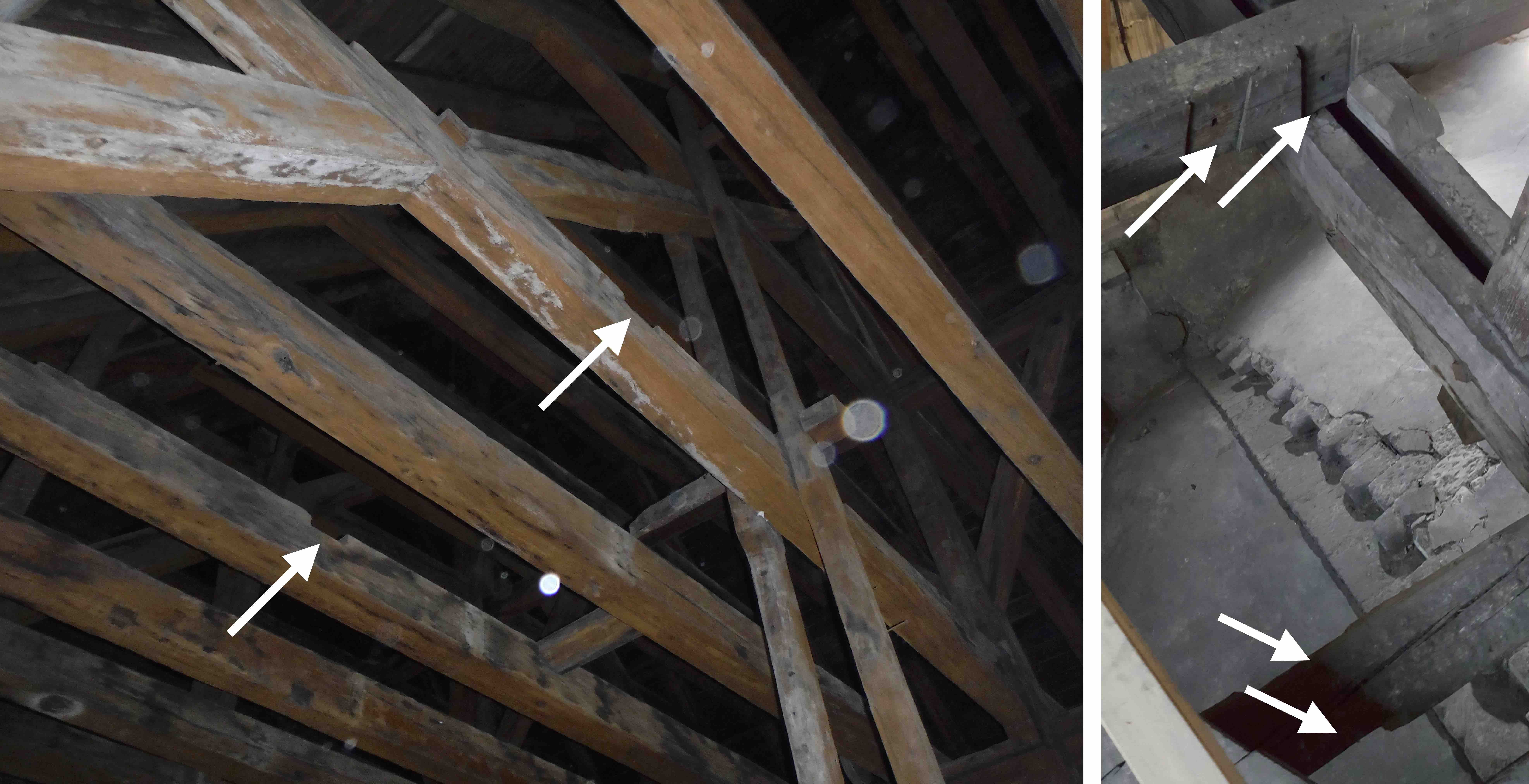}
\caption{Notches  indicating the reuse of ancient  wooden beams (by the author).}
\label{fig:2}
\end{center}
\end{figure}

The structures of the nave and choir used, at least in part, some timber beams of the more ancient, original roof of Notre-Dame, built around 1160-1170: several pieces of them, in fact, showed unused mortices or {\it mi-bois} notches  in some parts of  the charpente built after 1220, a clear sign of reuse, \cite{aubert1}, Fig. \ref{fig:2}. The reconstruction of the {\it charpente} was the consequence of a set of changes made on the cathedral during its construction. In particular, the guttering wall, i.e. the upper part of  the clerestory, was raised of about 2.70 m above its original height, a fact that had a number of consequences on the structure of the new {\it charpente}, as discussed below.


The {\it charpente} was a very large structure: about 115.6 m long, 13 m wide, 9.75 m high for the  nave and the choir. Though in \cite{dubu} it is said that the \cha\ was made with chestnut wood, it is certain that it was realized with the wood of oak trees. Scholars desagree about the quantity of trees employed for the structure: according to F. Épaud, who has deeply studied the timber structures of the Middle Ages, it was composed by the wood of about 1000 oaks, almost all of them with a diameter of $\sim25\div30$ cm  and 12 m high, a small part with a diameter of $\sim$50 cm and 15 m high, corresponding to about 3 hectares of forest. In an interview, \cite{reporterre}, A. Corvol, an expert in forests, evaluates at 3000 to 5000 m$^3$ the volume of the {\it charpente}'s wood, corresponding to about 2000 oaks; other estimations  give the value of 1300 oaks for 21 hectares of forest. The quantity of wood was, anyway, very important, so that the {\it charpente} was called {\it la forêt}, the forest: the visitor was immersed in an amazing very intricate set of wooden beams and struts, like in a sort of artificial forest crystallized during several centuries, Fig. \ref{fig:1}.
 
The {\it charpentes} of the nave and choir were different; built after that of the choir, the nave's one showed some changes in the design of its structure that let us think of an evolution in the constructive thinking of the ancient carpenters,  \cite{epaud1}, \cite{epaud2}. This point will be considered below in detail.

\begin{figure}
\begin{center}
\includegraphics[height=0.25\textheight]{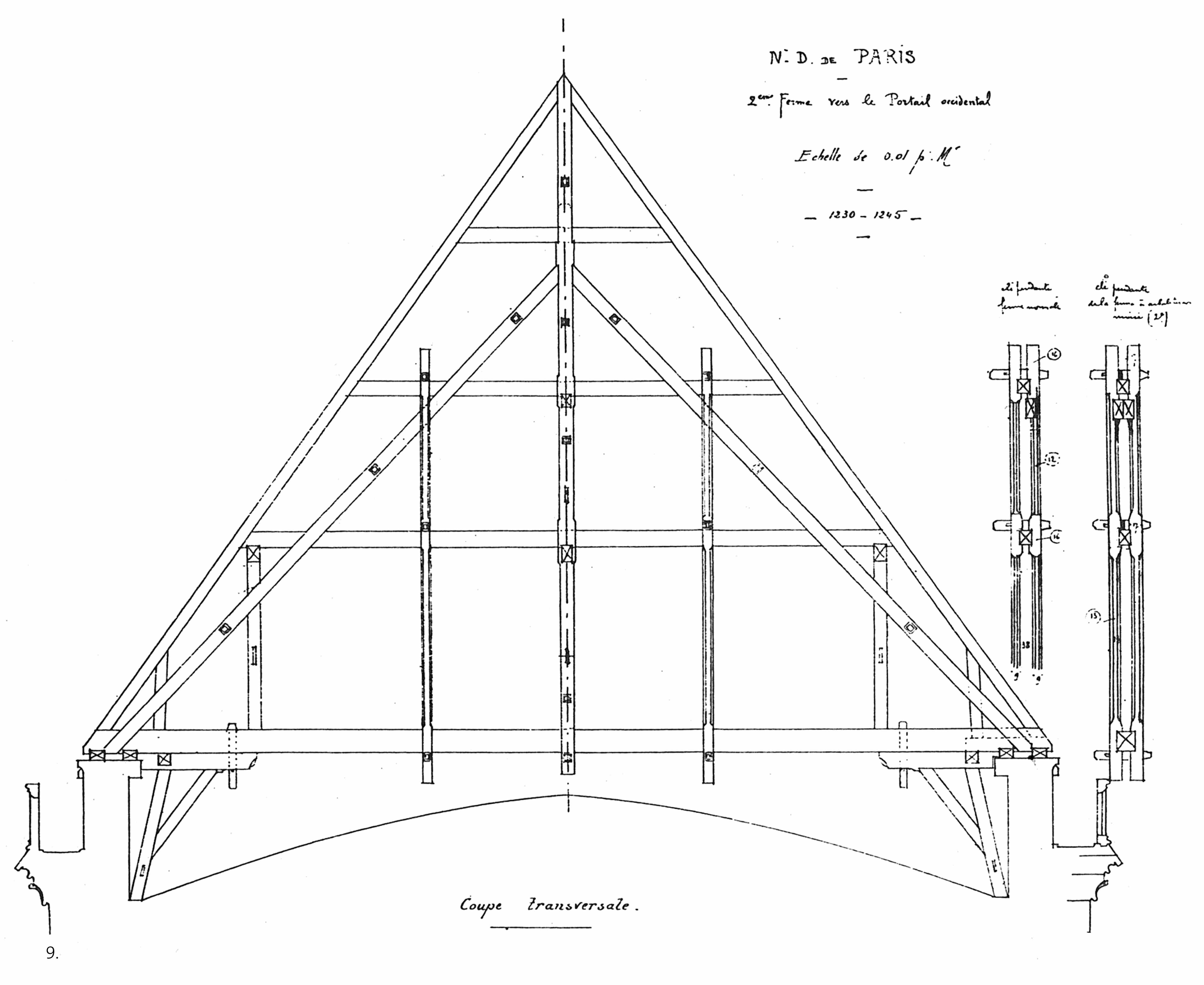}
\caption{The survey of the second {\it ferme}, FN2, of the nave, according to \cite{deneux}; the {\it suspente} on the right side was not present, in reality.}
\label{fig:3}
\end{center}
\end{figure}

The \com\ of Notre-Dame have been extensively studied in the past. A first relevant study of them  is that presented by Viollet-le-Duc for the  term \cha\ of his celebrated {\it Dictionnaire}, \cite{viollet}. Some basic concepts are introduced by the author and put into a historical perspective. However, his analysis remains qualitative and his hypotheses and concepts are still to be verified quantitatively. 
The structure of the \cha\ of Notre-Dame is also briefly considered in another well-known treatise on the history of architecture, that of Choisy, \cite{choisy}, Tome II. Choisy, like Viollet-le-Duc, gives a correct account of the evolution of the structural types of the roofing timber structures in France during the XIIth and XIIIth century, but his analysis of that of Notre-Dame is manifestly false: the \cha\ of the choir is not the one  schematically shown in his figure 6, page 328, and its differences with that of the nave are more numerous than those indicated by Choisy. 
Later, Deneux publishes a survey of the \cha\ of Notre -Dame, \cite{deneux}, in a work on the evolution of the {\it charpentes} from the XIth to the XIIIth century, see Fig. \ref{fig:3}.
In \cite{bechmann}, the author gives perhaps the  deepest analysis of the evolution of timber constructions for \com\ during the Middle Ages, the case of Notre-Dame is however only briefly cited.
More recent works are the MSc thesis of V. Chevrier, \cite{chevrier}, that contains a schematic  survey of the \cha, and which is, essentially, a first study on the dendrochronology of the wood of the structure, and \cite{fromont}, that also contains 
a survey and a description of the \cha. In \cite{sandrontallon} the history of the construction of Notre-Dame is retraced, but somewhat curiously almost without any consideration about the \com, while in \cite{hinslop} the \cha\ is just described using the figures of the {\it Dictionnaire} of Viollet-le-Duc.
Finally, the already cited works of F. Épaud, \cite{epaud1,epaud2} give a general description of the timber structure of Notre-Dame, particularly form the  point of view of the history of timber constructions in the Middle Ages.

All of these works are rather descriptive, they tend to give a historical perspective of the Notre-Dame's \cha\ in the context of the technical evolution of its epoch or to analyse some peculiar aspects, like those linked to the use and exploitation of forests in the France of the Middle Ages. However, as  confirmed in \cite{epaud2}, there is a lack of structural analysis for the \com\ of Notre-Dame. This paper is a first attempt to fill the gap. The objective is twofold: on the one hand, this study aims at giving an evaluation of the structure of the ancient \cha, to assess its condition before the fire, from a mechanical point of view, and to help in this way in taking decisions about its reconstruction. On the other hand, it aims at going beyond the mere descriptive analyses done so far and to try, by a precise structural study, to shed a light on its real static behavior, how, presumably, it was thought by its ancient masterbuilders. In some way, it is an attempt to retrace the constructive thinking of the Gothics, their ideas in designing their {\it charpentes} and to check whether  some of the more common ideas on this matter are sound or not. In this paper, only the \cha\ of the choir and of the nave are studied, not that of the transept, rebuilt after 1843 according to structural schemes of the period.


\section{Brief historical account}
The objective of this Section is just to briefly recall the main historical facts concerning the transformations of the \com\ of Notre-Dame, as they have strongly influenced  the structural functioning of the \cha. The first roofing structure of the cathedral, on the choir, was undoubtedly finished   before 1182 (\cite{aubert1}, page 16, \cite{bruzelius87}), when the choir was consecrated. Subsequently, a new \cha\ was erected, as said using, at least in part, the elements of the ancient one: on the choir between 1225 and 1230 and in the nave between 1230 and 1240, \cite{aubert2}.
The reasons for this reconstruction are not well known and historians still debate about that. According to \cite{aubert1}, a fire was set to the cathedral during the night of the Assumption 1218, by a thief trying to steal some precious {\it candelabra} (\cite{sandrontallon} specifies that the thief was an english man), what could, at least chronologically, justify the reconstruction of the \com\ from 1220 on. Whether or not this fire really took place, the modifications done to the still unfinished cathedral starting from the second decade of the XIIIth century are important, they do not concern uniquely the \cha\ and they are still today a matter of historical debate, cf. \cite{mark84,bruzelius87,murray,sandrontallon}.

We cannot  state here with certitude about the reasons for the architectural changes of the cathedral (just a matter of style, like in \cite{sandrontallon}? Also, at least in part, a matter of structural response, according to \cite{mark84}?), but such transformations necessarily forced the carpenters to adopt a static scheme different from and better than the previous one. 
In \cite{viollet}, at the term \cha, the author gives a personal explanation of this transformation, actually a general change in the construction of the Gothic {\it charpentes}, according to him put in place for the first time exactly with the new \com\ of Notre-Dame. 

 \begin{figure}
\begin{center}
\includegraphics[height=0.23\textheight]{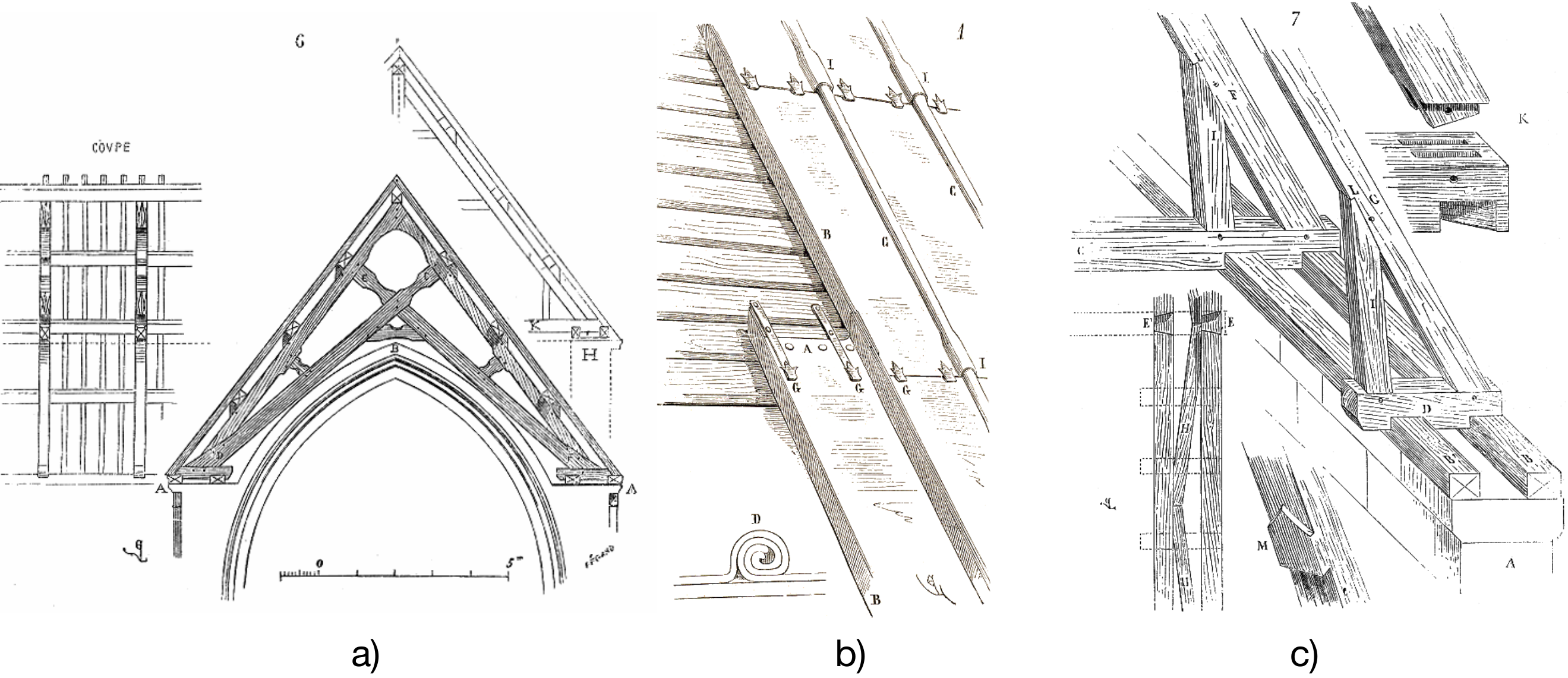}
\caption{Structural schemes from \cite{viollet}: a) scheme of a {\it ferme} without {\it entrait}; the \cha\ of $\sim$1170 was probably similar to this one; b) the system of lead tiles observed by Viollet-le-Duc on the roof of the cathedral of Chartres, before the fire of 1836 that destroyed entirely the \com; it is this system that he used to cover the roof of Notre-Dame of Paris after the restoration of the mid of the XIXth century; c) the system for the support of the {\it fermettes} (G), composed by {\it blochets} (D) and {\it sablières} (B), in a \cha\ with {\it chevrons formant ferme}.}
\label{fig:4}
\end{center}
\end{figure}

The crucial architectural transformation of the cathedral operated after 1220 that with no doubts is directly linked to the change of the structural type of the \cha\ was the raise of the guttering wall, already mentioned above. This allowed the carpenters to use frames ({\it fermes} in French) with an {\it entrait}, i.e. a long tie-beam relying the bases of the frame's rafters (the {\it chevrons}, in French), which is the ideal device to absorb the horizontal component of the force in the rafters. In the previous roofing structure of $\sim$1170, the vault exceeded in height the top of the guttering wall, so it was not possible to use {\it fermes} with {\it entrait}. We do not know how this older \cha\ was made, we can just imagine that it was something like that in Fig. \ref{fig:4} a). This type of solution was practically compulsory for vaults having the keystone higher than that of the {\it arcs formerets}, i.e. of the longitudinal ribs of the vault, those belonging to the wall of the clerestory, how it was common in France in the XIIth century, \cite {sandrontallon}, page 62, and practically the rule with sexpartite vaults, like those of Notre-Dame. 

\begin{figure}
\begin{center}
\includegraphics[height=0.23\textheight]{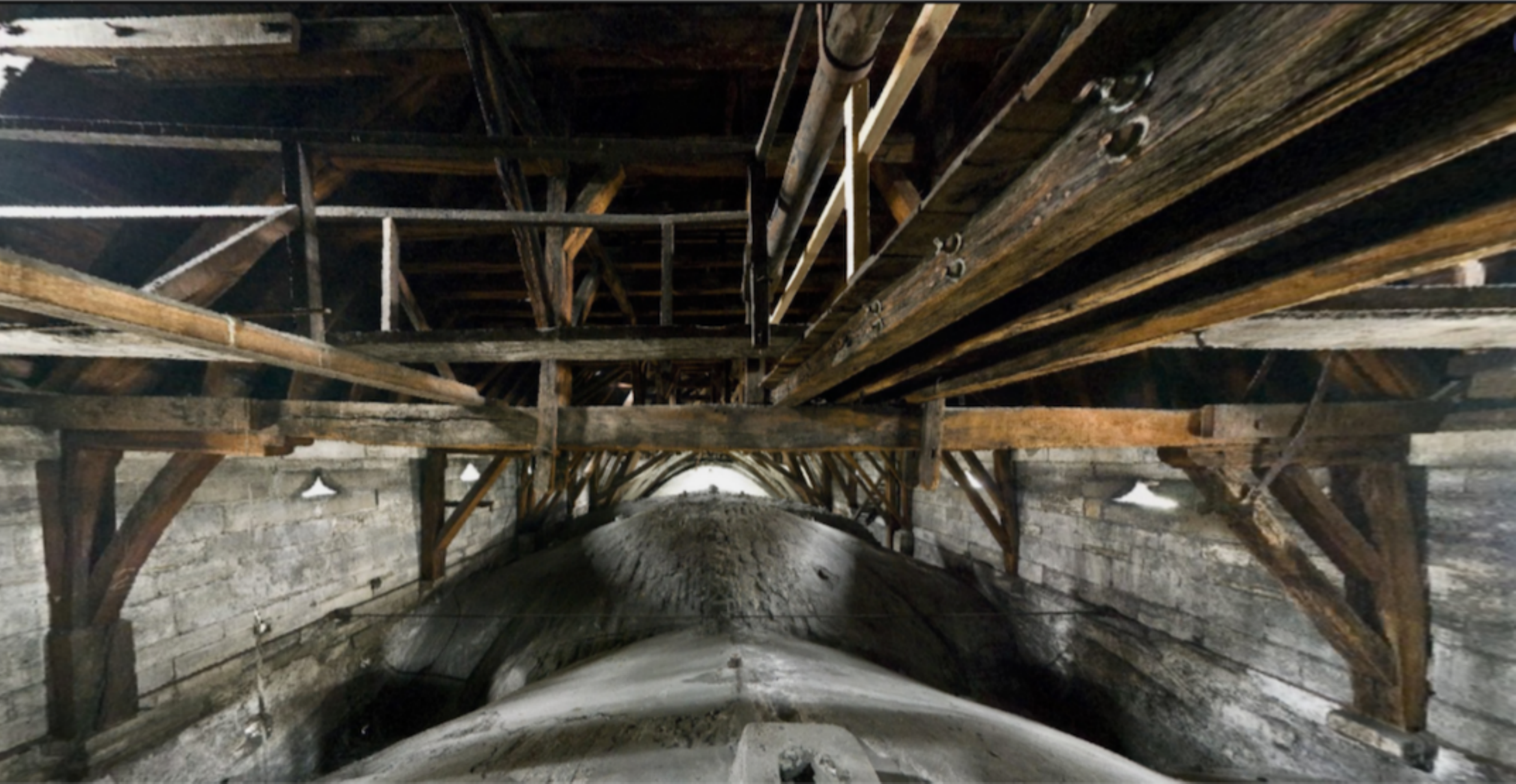}
\caption{The \cha\ of the choir; the top of the guttering wall was, before $\sim$1220, at the level of the corbels supporting the wall pieces and braces beneath the {\it fermes} (extracted from \cite{laserscan}).}
\label{fig:5}
\end{center}
\end{figure}

We have already mentioned that during the modifications started around 1220, the guttering wall was raised of $\sim$2.70 m above the keystone of the high vault, see Fig. \ref{fig:5}, where it is apparent that the {\it entraits} of the {\it fermes} can now well pass above the vault, while this was impossible for the \cha\ of $\sim$1170, as the top of the guttering wall was at that time at the level of the stone corbels well visible in the picture and serving as support for the lower part of the {\it fermes}. 

Different hypotheses can be made about these changes. In the case the raise of the guttering wall was dictated solely by aesthetic considerations, the carpenters had the possibility to change the structural scheme of the \cha, passing to a more efficient one. In fact, as pointed out in \cite{viollet}, the solution depicted in Fig. \ref{fig:4} a) was submitted to deformations of the entire structure, caused mainly by the bending of the rasters, that determined mouvements of the footholds, not restrained by a tie like the {\it entrait}. The masterbuilders were certainly aware of this fact and did not hesitate to change the scheme of the main frames. Anyway, another hypothesis can be made: it is possible that the raise of the guttering wall was needed because the ancient \cha\ manifested the structural problems described above and the carpenters suggested to modify the clerestory, raising the guttering walls to adopt the scheme of {\it fermes} that they actually used after 1220. In other words, we can think that it was the need for a change of the roofing structure to impose the raise of the guttering wall. Of course, nobody can affirm this fact with certitude but it cannot be excluded. We could have an indirect proof, by calculation, of this hypothesis if we knew how the ancient \cha\ of the XIIth century was made, which is not the case, unfortunately.

Whartever the reason for the raise of the guttering wall, what is certain is that the carpenters of the XIIIth century were facilitated in the design of their \cha\ but they were also with no doubts aware of some new problems. The first one, is just that of the {\it entraits}, the second one, that of wind. 
The use of long {\it entraits} (at Notre-Dame, they have a length of $\sim$13 m), needs to dispose of beams of great section, to withstand the bending of the rod. Such a type of beams very probably did not exist in the ancient \cha\ of $\sim$1170, because of a completely different static scheme of the frames. So, the builders of the XIIIth century were faced to the problem of finding a sufficient number of oaks with large diameter (\cite{epaud1} estimates trunks of $\sim$50 cm of diameter and $\sim$15 long). This was not so easy during the XIIIth century (it is famous the adventurous pilgrimage of Abbot Suger to find trees of sufficient size for the roof of the Royal Abbey of Saint Denis), and the use of a \cha\ with few {\it entraits} was hence almost compulsory. The carpenters of the epoch adopted a solution that is known in French as the \cha\ with {\it chevrons formant ferme}. Such an original construction had different advantages, in \cite{viollet} the author cites some of them beyond the one suggested here. According to the historians of architecture, they were mainly two: the possibility of using, excluded the {\it entraits}, pieces of wood of relatively small dimensions (say, trunks of $\sim25\div30$ cm of diameter), that has also the advantage of obtaining relatively light structures, and the distribution almost continuous of the loads of the \cha\ on the guttering wall. 
Some historians, e.g. \cite{enlart}, cited in \cite{bechmann}, have pointed out here a discrepancy with the structural doctrine of Gothic architecture: on one side, the structure of a Gothic cathedral is arranged so as to carry the loads at some points (the pillars, columns, buttresses), i.e. it is organized as an array of isolated points of force.
On the other hand, the carpenters, working on the same building at the same period, went in the opposite direction: they conceived structures that distributed more or less continuously the load on the upper part of the clerestory wall. 

This point of view, we will see, is rather illusory: numerical simulation shows that, actually, the loads transmitted by the \cha\ to the guttering wall are far from being uniformly distributed and the reason of that is exactly the global structural organization that the carpenters gave to  the \cha, i.e. its three-dimensional functioning. We will show  this fact in  Sect. \ref{sec:reactions}. What is sure, anyway, is  that carpenters did not care about a correspondence of the main frames of the \cha, the {\it fermes principales}, carrying the most part of the load, and the pillars or the flying buttresses of the cathedral: actually, they do not correspond at all. For instance,  the nave is divided into seven sexpartite vaults, while the \cha\ into thirteen parts; for the choir, there are five vaults while the \cha\ is divided into eight parts. Hence, there is nowhere a correspondence between the pillars supporting the vaults or the flying buttresses and the main frames, that transmit the most part of the load to the stone structure. The only fact that we can affirm is that the carpenters of the XIIIth century did not take into any consideration this correspondence of load-points for the \cha\ and the stone structure beneath. In \cite{enlart} this discrepancy is severely criticized, and the criticism is based, on one hand, upon a rather ideological point of view, while, on the other hand, on a static idea that cannot be considered as valid. Actually, the carpenters of Notre-Dame did not have the choice: they simply could not realize, with the beams at their disposal, a \cha\ whose main frames were spaced of  the same distance between the pillars, which is of $\sim$5.6 m. Actually, in the choir the distance between the {\it fermes principales} is of 4.1 m, reduced to 3.5 m in the later nave \cha, a change that has been, probably, suggested to the builders  by wariness, after the construction of the choir \cha.

The problem of the wind is less considered in the literature, but it was with no doubts present to the mind of the builders, as some details of the \cha\ suggest.  In fact, it is evident that the raise of the guttering wall implies problems of equilibrium for this last when the roof is submitted to the action of the wind. It is likely that the builders of the XIIIth century were aware of that and they certainly grasped the need for a vertical load for the wall to withstand the horizontal forces of wind. The lightness of the \cha\ was hence not at all dictated by mechanical reasons, as suggested by several authors, namely by Viollet-le-Duc, but more likely by economical reasons (the less the material, the less the price) and probably by the penury of oak trunks of large dimensions, \cite{bechmann}. That is why the carpenters of the new \com\ of Notre-Dame  had the need to invent an effective system for transmitting to  a so high guttering wall the horizontal thrust of the wind acting on the roof. This point is crucial to try to understand the constructive thinking of the builders of the XIIIth century \cha, and it is considered in depth in Sect. \ref{sec:horizforce}.

\begin{figure}
\begin{center}
\includegraphics[width=\textwidth]{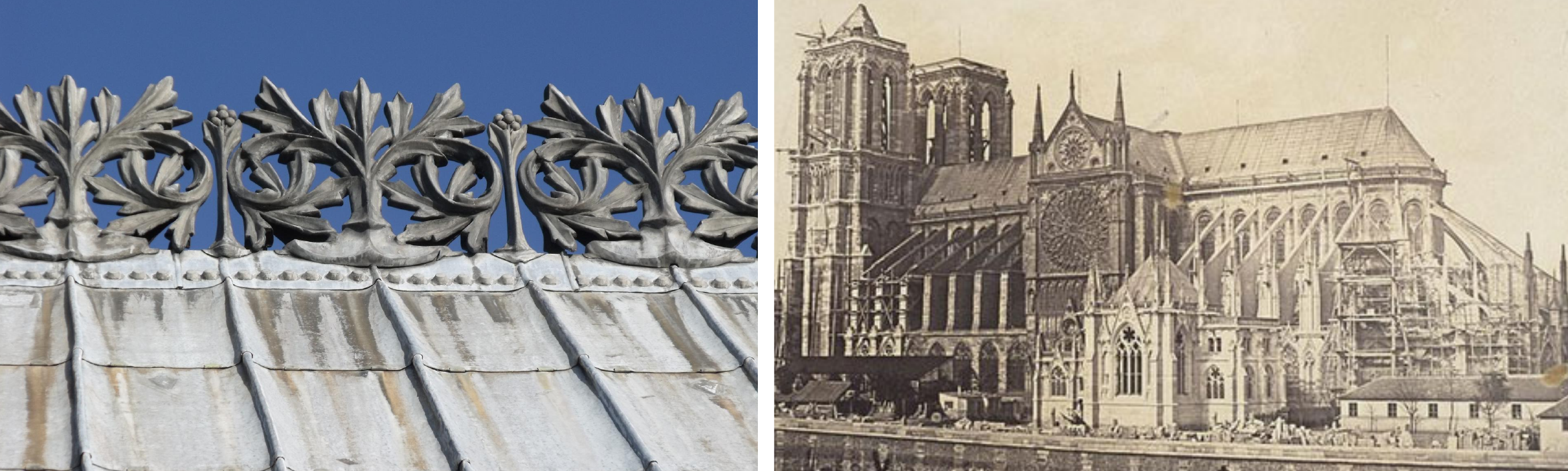}
\caption{The lead frieze at the roof top and Notre-Dame before the restoration of Lassus and Viollet-le-Duc, without the frieze.}
\label{fig:7}
\end{center}
\end{figure}

Originally, according to \cite{dubu}, who wrote a history of Notre-Dame during the restoration works of Lassus and Viollet-le-Duc,  the covering of the roof was made by 1236 lead plates, each one of the dimensions 0.975$\times$3.248 m and 5 mm thick, for a mass of 120 kg each, which gives a total mass of the lead covering of $\sim$2060 tons. This calculation is manifestly false: if the dimensions of a tile are correct, its mass should be of $\sim 180$ kg, and the total mass of the covering of 221.15 tons, not far from the value normally given of 210 tons. 
However, this was not the lead covering melt by the fire of April 15th, 2019. In fact, in \cite{lassusviollet} it is specified that the Cardinal of Noailles proceeded to the restoration of the roof in 1726 and that the new mass of the lead covering was of 220240 pounds, i.e. about 99.9 tons. Evidently, the new lead tiles had a less thickness, presumably of $\sim$2.25 mm.
Successively, during the restoration of Lassus and Viollet-le-Duc, the covering was made again, still with lead tiles, but now with a thickness of 2.82 mm, \cite{laboulaye}; the tiles were welded together and fixed to the \cha\ with staples, Fig. \ref{fig:4} b).  This was the lead covering  melt by the fire of 2019. Its total mass can be estimated to $\sim$135 tons. Also, a lead frieze was put on the top of the roof, that did not exist before, Fig. \ref{fig:7}.

This succinct  report put in evidence the strong relation existing between the historical vicissitudes of the cathedral and the construction and transformations of the \com. This should be bear in mind for any interpretation of a structural analysis done on the \cha. We will reconsider these aspects in the following of the paper. It is important now to describe in detail the structure of the \com, what is done in the next Section.

\section{The structure of the \textit{combles}}


A \cha\ with {\it chevrons formant ferme} is an articulated structural system, thought with a main objective: to realize a structure employing wooden beams, ties and struts that can be obtained from trees of ordinary dimensions, say trunks of a diameter of $\sim20\div30$ cm. All is conceived with this objective. In this Section, we first consider in detail the geometry of the \com\ and then we analyse the  mechanical functioning of the \cha.

\subsection{Description of the Notre-Dame \textit{charpente}}
The structure of \com\ with {\it chevrons formant fermes} is composed of two main parts: the {\it fermes principales} or {\it chevrons maîtres}, the main frames, and the secondary frames, in French the {\it fermes secondaires} or {\it fermettes} or {\it chevrons}. The \cha\ is assembled by the repetition of structural units, each one composed by a main frame and by a certain number of secondary frames. These last are put at a distance varying between $\sim$60 and $\sim$80 cm, i.e., they are rather close together. Typically, a structural unit makes use of 4 to 5 {\it fermettes}, so it has ordinarily a length of $\sim3\div4$ m. In the choir of Notre-Dame the main frames are spaced of $\sim$4.1 m and four {\it fermettes} are used, spaced of 82 cm. In the nave, this distance has been reduced to $\sim$3.5 m and the {\it fermettes}, still in the number of four, are spaced of 75 cm. The main frame is then linked to the {\it fermettes} by a longitudinal structure, composed by horizontal beams, the {\it liernes} (we adopt here the same jargon used in \cite{chevrier}, though normally the term  {\it lierne} is used to indicate  a rib) and by reinforcing struts, the {\it aisseliers} or {\it liens} (braces). 

\begin{figure}
\begin{center}
\includegraphics[width=\textwidth]{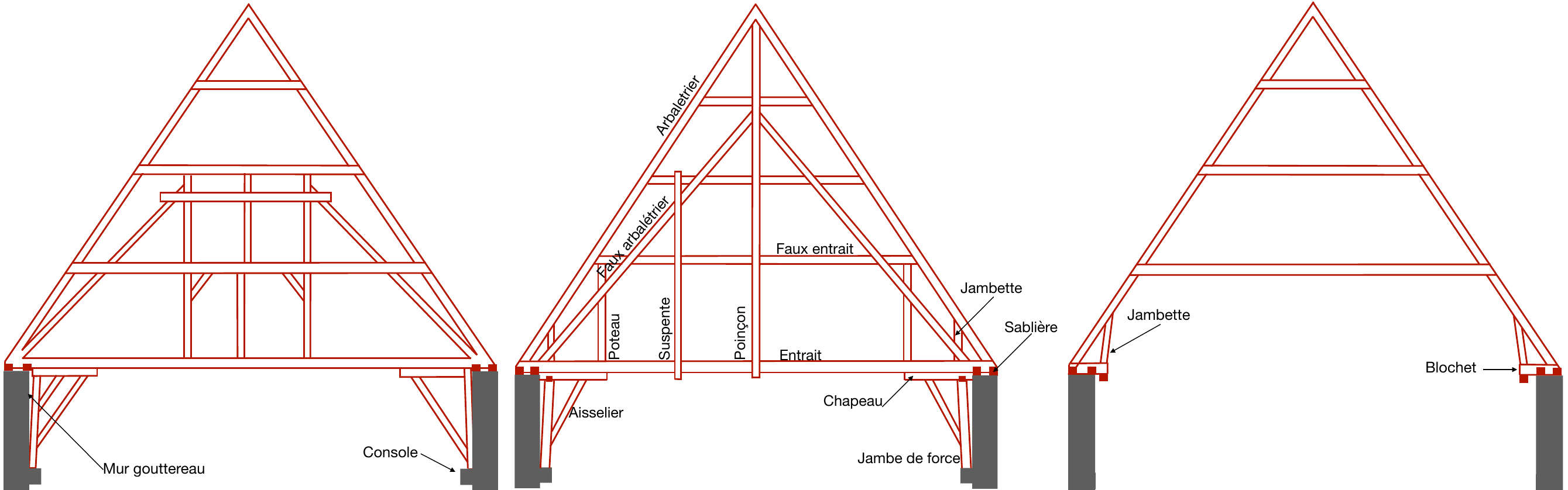}
\caption{From left to right, scheme of a {\it chevron maître} of the choir, of the nave and of a  {\it fermette}.}
\label{fig:8}
\end{center}
\end{figure}

The {\it chevrons maîtres} of the choir and nave of Notre-Dame, as well as the model of a {\it fermette}, are shown in Fig. \ref{fig:8}. In the same figure, the French names for the different pieces of the structure are also indicated, they will be used in the remainder of this paper. 
 These schemes have been reconstructed using mainly \cite{chevrier} and \cite{fromont}.  These documents give a lot of valuable information about the structure of the \com, namely on the dimensions of the wood pieces.

\begin{figure}
\begin{center}
\includegraphics[width=\textwidth]{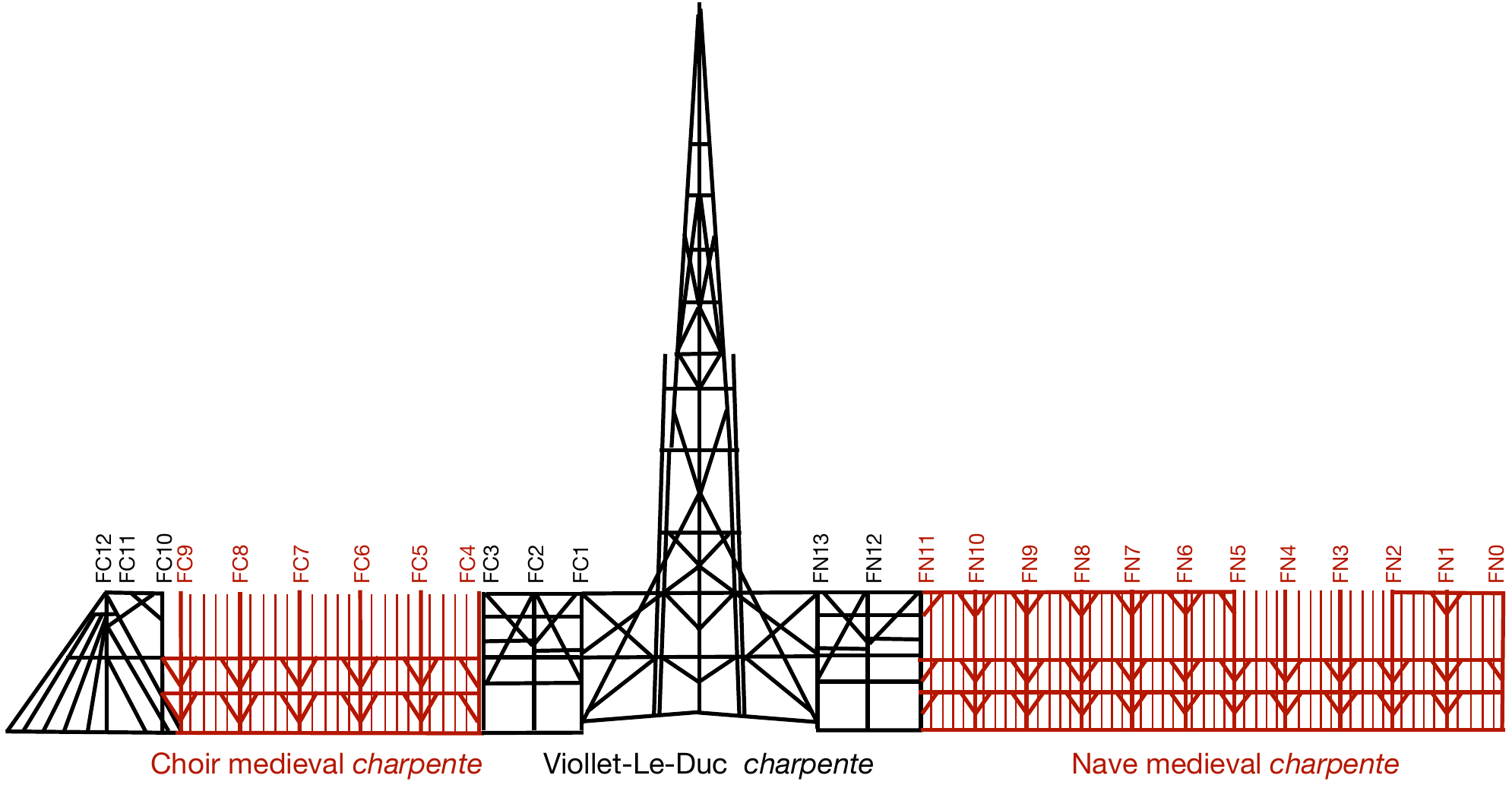}
\caption{Longitudinal scheme of the \cha; the parts of the structure studied in this paper are in red; thick lines indicate the main frame and the longitudinal bracing system, while thin lines represent the secondary frames (scheme reconstructed from the survey in \cite{fromont}).}
\label{fig:11}
\end{center}
\end{figure}

The longitudinal scheme of the entire \com\ is presented in Fig. \ref{fig:11}. The three parts of the \cha\ are clearly indicated as well as the notation of the the {\it fermes principales}, as usually adopted. The longitudinal structure relying together the {\it fermes principales} and the {\it fermettes}, formed by {\it liernes} and {\it aisseliers} is schematically presented too.  In this paper, only the parts of the \cha\ between {\it fermes}  FC4 and FC9, for the choir, and between FN1 and FN11, for the nave, are studied. They correspond to the regular part of the medieval \cha, the rest being the part of the structure constituting the apse of the choir or that reconstructed by Lassus and Viollet-le-Duc during the XIXth century, and based upon a different structural scheme, that of {\it fermes, pannes et chevrons}, \cite{bechmann,bernardi}.
Above the {\it chevrons} of the main or secondary {\it fermes}, there is a wooden decking, the {\it voligeage}, this last supporting the lead tiles, nailed on it, see Fig. \ref{fig:4} a). 

While the choir's \cha\ was rather homogeneous from FC4 to FC9, this was not the case for that of the nave. A first difference between the parts of the nave's \cha\ was the  fact that the longitudinal beam put on the top of the {\it chevrons}, the so-called {\it panne faîtière}, was present, with its  {\it aisseliers},  only between FN0 and FN2 and then between FN5 and FN11, while it was absent in the rest of the structure, between FN2 and FN5 (in the choir, it is absent everywhere between FC4 and FC9). This part of the structure was a part of the bracing system of the \com\ and its role was important especially during the construction phase. Its presence in the nave's \cha\ shows that the carpenters, after the experience of erecting the choir's structure, had understood its importance for the global equilibrium of the \cha. However, its absence between FN2 and FN5 cannot be easily explained. May be, it is linked to the absence, in FN3, of the upper part of the {\it poinçon}, the vertical tie relying the {\it entrait} to the top of the frame,  Fig. \ref{fig:17} d). This could be the fact of a local rupture, not followed by reparation.
  \begin{figure}
\begin{center}
\includegraphics[width=.3\textwidth]{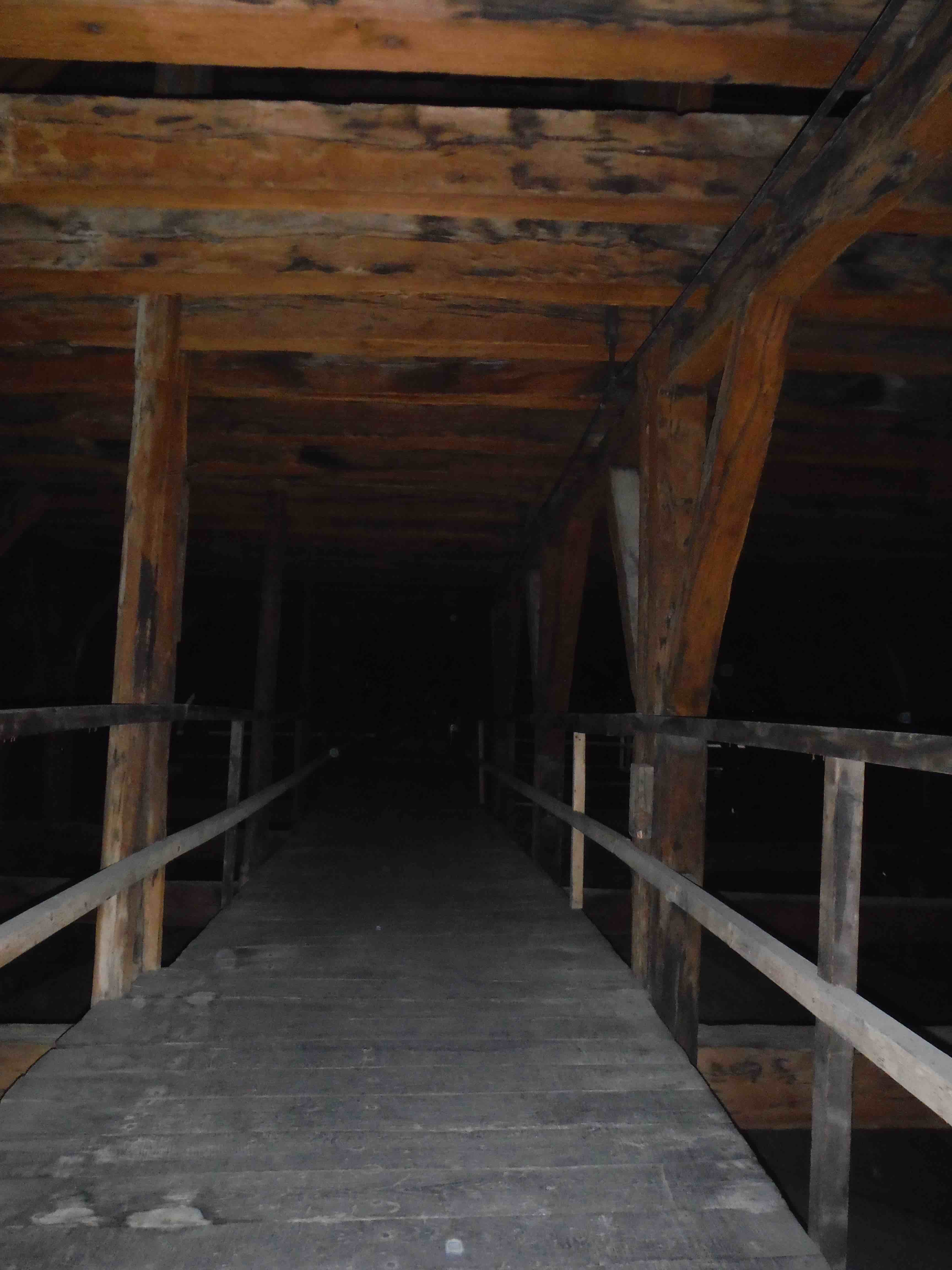}
\caption{The catwalk; the {\it suspente} is the vertical tie on the left, while the {\it poinçon} the one on the right, with its {\it aisseliers} (by the author).}
\label{fig:12}
\end{center}
\end{figure}

\begin{figure}
\begin{center}
\includegraphics[width=\textwidth]{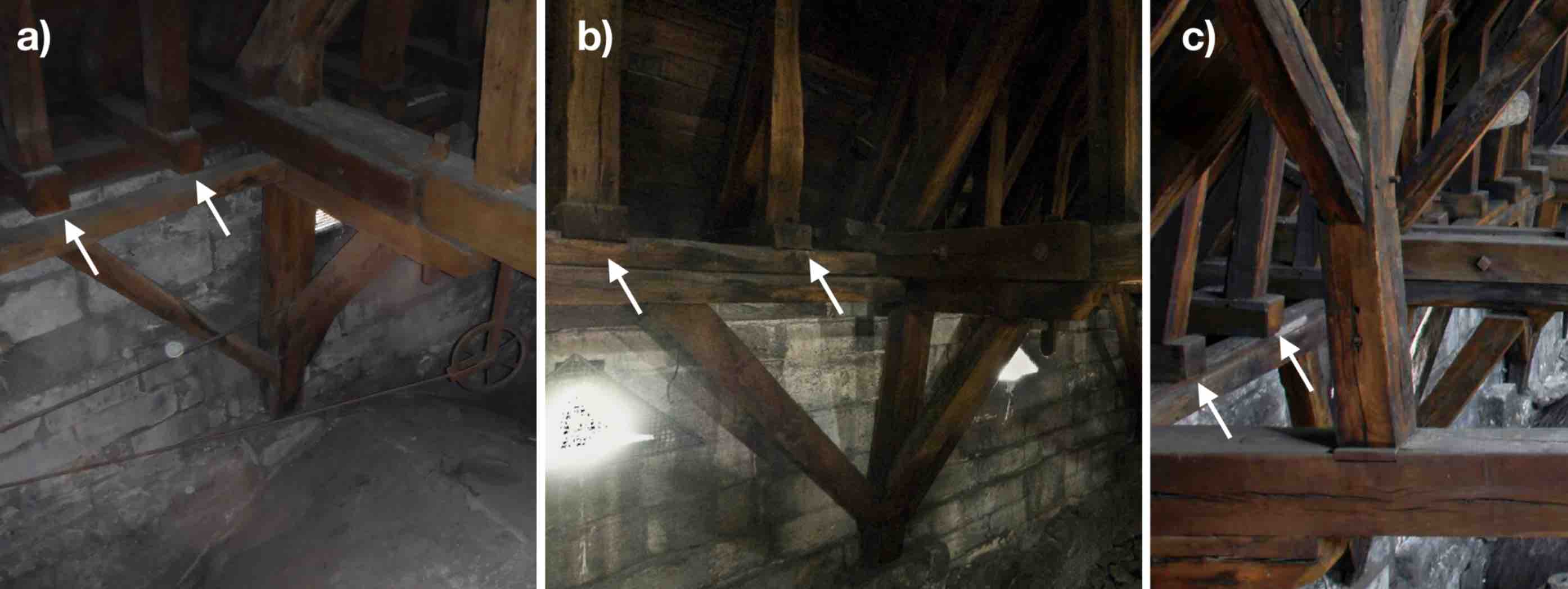}
\caption{The detail of the contact between the {\it blochets} and the longitudinal tie, in the nave's \cha: a) the lack of contact  for the part close to FN3, North side (by the author); b) the contact ensured by a thick piece in other parts of the \cha\ (extracted from \cite{laserscan}); c) the direct contact close to FN6, South side (from \cite{fromont}).}
\label{fig:13}
\end{center}
\end{figure}

Another point of debate is the presence of the {\it suspentes} on the North side of the nave's \cha. As apparent from Figs. \ref{fig:8} and \ref{fig:17} c), this was a clear element of dissymmetry for the frame. In \cite{fromont} a plausible hypothesis for this dissymmetric solution is given: the {\it suspente} was used to support the {\it entrait} where a catwalk, running all along the \com, was placed, between the {\it poinçon} and the {\it suspente}, Fig. \ref{fig:12}. This hypothesis, that seems to be confirmed by dendrochronological analyses, is the only reasonable one, from the structural point of view. Anyway, in correspondence of FN3 at least, the link between the {\it suspente} and the {\it entrait} was broken. Still in correspondence of FN3, on the North side, there was not contact between the {\it blochets} and the longitudinal tie running between the {\it jambes} of the {\it chevrons maîtres}, see Fig. \ref{fig:13} a). However, this was not the rule: in other parts, perhaps everywhere else, this contact was ensured, see Fig. \ref{fig:13} b). The reason for this lack of contact, that is undoubtedly a defect from the structural point of view, is argued in \cite{fromont} as the probable use of the longitudinal tie only for the constructive phase. This is not likely: all the {\it blochets} of the nave's \cha, unlike those of the choir, jut outside the thickness of the guttering wall, and were charged by the {\it jambettes} exactly above the longitudinal tie, Fig. \ref{fig:13} b). There cannot be any doubt that the longitudinal tie was  put in place to relieve the bending of the {\it blochets} and better support the {\it jambettes}, which is confirmed by the contact that actually existed between the {\it blochets} and the tie in other parts of the \cha, contact at least in some parts ensured through a thick piece, like in Fig. \ref{fig:13} b). This is a clear sign that the builders wanted to use the longitudinal tie not particularly (but how?) in the constructive phases, but principally for the structural functioning of the whole \cha. The lack of contact discussed here can be imputed, very probably, to the absence of a thick piece between the {\it blochets} ant the longitudinal tie. It is likely that this thick piece existed at the beginning and that, afterwards, it disappeared for some reason, e.g. it was fallen for the inescapable and physiological mouvements of the timber structure and never put back in its place for a lack of maintenance during the ages. 
The discrepancies between the structure close to FN3 and the rest of the nave's \cha\ cause some differences in the structural response of the \com\ that have been considered in the numerical simulations  on the structural model of the nave's \cha\ as special cases, see Sect. \ref{sec:structanalysis}.

The two cited documents, \cite{chevrier} and especially \cite{fromont}, have been the main source for constructing the mechanical model described below. There are anyway some more precise documents: the first one, is the 3D laserscan survey made by A. Tallon in 2010, a pioneer of this technology applied to Gothic cathedrals, \cite{tallon}. This rather complete survey of Notre-Dame is available on the site {\it TruView Notre-Dame}, \cite{laserscan}, and it comprehends a view of the choir's \com\ and another of the nave's \cha. Unfortunately, these views are only partial and also if the web application allows to take measures on the interactive survey, it is almost impossible to obtain reliable dimensions of the wood beams. However, the few measures that it has been possible to obtain reliably are close to  those  extrapolated from \cite{chevrier} and \cite{fromont}.
The second document is still a laserscan 3D survey, made in 2014 by the firm {\it Art Graphique et Patrimoine}, but this document is not at the disposal of the scientific community. The documents used for the reconstruction  of the \cha\ are anyway sufficiently detailed to obtain, on the whole,  a reliable mechanical model of the structure. In fact, though the \cha\ was constituted by repetitive elements, the {\it chevrons maîtres}, the {\it fermettes} and the {\it liernes}, these were not perfectly identical. The most part of the wooden trunks had been modeled by hand, squared with an ax; their dimensions were similar but not perfectly identical, their thickness often not constant throughout the axis, several geometrical imperfections were present everywhere in the structure. As a consequence, it is impossible, in practice, to obtain a mechanical model with exact, precise dimensions, a certain degree of approximation, say of the order of the centimeter, must be accepted. This is normal for any civil structure, and also for a so ancient  and huge timber structure, not realized with the modern industrial tools but handcrafted. We will see in Sect. \ref{sec:structanalysis} that, anyway, the degree of precision allowed by the documents used for the reconstruction of the \cha\ is well sufficient.

\subsection{Mechanical functioning of the \textit{charpente}}

We ponder now on the structural functioning of a \cha\ with {\it chevrons formant ferme}. In the literature, some qualitative and partial explanations on this matter can be found, e.g. in \cite{viollet} or in  \cite{bechmann}. Let us  consider them in a global view on the structural behavior of this kind of structures, that will then be corroborated by the numerical simulations presented in Sect. \ref{sec:structanalysis}. 

  The main difference between the {\it ferme principale} and the {\it fermettes} is the presence of the {\it entrait}, used only for the main frame. The horizontal thrust at the base of each {\it fermette} needs hence to be equilibrated in another way, otherwise the bending of the {\it chevrons} or {\it arbalétriers} would severely deform the {\it fermette}. The way invented by the Gothic carpenters is that of the {\it sablières} and {\it blochets}, see Fig. \ref{fig:4} c). Each  {\it chevron} of a  {\it fermette} has at its base a {\it blochet}, hosting the {\it chevron} and a {\it jambette}. The {\it blochet} is not directly placed on the top of the guttering wall, but it is linked to and supported by two longitudinal rods, the {\it sablières}. These two wooden  beams are restrained by the {\it entraits} of two successive {\it chevrons maîtres}. In this way, the horizontal thrust at the base of each {\it fermette} is absorbed by the system of the {\it sablières}. These ones are hence solicited in bending and shear in the horizontal plane and transfer in this way the resultant of the horizontal thrust of the {\it fermettes} between two {\it chevrons maîtres} to the {\it entraits } of these last.  The whole system hence does not apply any horizontal thrust on the top of the guttering wall. This is a fundamental fact of the \cha$s$ with {\it chevrons formant ferme}: all the system was designed to transmit only vertical forces to the top of the guttering wall, while the horizontal forces engendered by the vertical loads, i.e. by the own weight of the structure, were self-equilibrated by the system composed by the {\it chevrons}, {\it blochets}, {\it sablières} and {\it entraits} of the main frames. Because the \cha\ is simply posed onto the guttering wall, the only mechanism able to absorb horizontal forces should be the friction  between the wood of the \cha\ and the stone of the wall. We will see in Sect. \ref{sec:horizforce} that actually this is not possible: not only friction cannot absorb the horizontal thrust of the Notre-Dame's \cha, but more importantly, such a situation could bring the guttering wall to a failure. In fact, the horizontal thrust applied to a high guttering wall could produce a rotation of the wall at its base, i.e. where it meets the vault, causing its collapse. It is likely that the builders of Notre-Dame were conscient of this danger, consequence of the elevation of the guttering wall, and adopted intentionally the system described above in order to create a structure not pushing horizontally on the wall's top.

  These considerations are, at least in part, given also in \cite{viollet}, with some differences. In fact, Viollet-le-Duc implicitly admits the role played by the system {\it blochets}-{\it sablières}, but he imputes its adoption to another reason. According to him, the reduction in dimensions of the Gothic cathedrals, with respect to the  Romanesque architecture, pushed the builders to use more and more thin walls, to a point that to pose the ancient system of {\it charpentes} with {\it fermes, pannes} and {\it chevrons} on the top of so narrow walls became problematic. Viollet-le-Duc argues that this lead the carpenters of the Gothic age to introduce the \cha$s$ with {\it chevrons formant fermes}, having a less global thickness, and to increase the slope of the roof, in Notre-Dame it is  $\sim55^\circ$. Rather surprisingly, Viollet-le-Duc does not make any structural consideration about the global functioning of the \cha; he just observes,  correctly, that, on one hand, the increase of the slope decreases the bending of the {\it chevrons}, so allowing the use of wood beams of relatively small cross sections, and, on the other hand, he takes care of the possible excessive bending of the two {\it sablières} between two consecutive {\it fermes principales}. He suggests that in some cases the carpenters placed some diagonal struts and ties between them, in order to create a horizontal truss, Fig. \ref{fig:4} c), detail (H). This implies that the Gothics had grasped, at least in an intuitive, experimental way, the mechanical functioning of a truss; here, we cannot affirm this with certitude. In fact, though  diagonal struts (the {\it écharpes}, in the French jargon) were common in timber constructions, their role was that of conferring  stability to the construction. When employed in a truss, their functioning is different, basically they increase the bending stiffness and strength of the structure. However, in Notre-Dame, such a device was not present. It seems hence probable that Viollet-le-Duc had considered the impossibility, and danger, of the transmission of the horizontal thrust from the bases of the {\it chevrons} to the top of the guttering walls  by friction. 
  
In \cite{choisy}, the point of view is different: according to Choisy, in fact, the increase of the roofs' slope in the Gothic period is essentially used to decrease the horizontal thrust. This interpretation cannot be considered as correct, for what said above. In the same way, the point of view of Pol Abraham, \cite {polabram}, as reported in \cite{bechmann}, is mechanically wrong: he considers, in fact, that to explain the deformations of the Gothic vaults, the horizontal thrusts of the above \cha$s$ should be put into consideration. In other words,  he implicitly admits  that there is a transfer, by friction, of the horizontal thrust of the \cha\ to the top of the guttering walls.  Actually, we will see in Sect. \ref{sec:structanalysis} that, on one had, the timber structures invented by the Gothics did not apply any horizontal thrust to the walls and, on the other hand, that such a thrust, if existing, had caused the failure of the guttering wall above the vaults. The point of view of \cite{bechmann} is not clear on this subject. If he strongly supports the idea of Viollet-le-Duc of the increase of the roofs' slope to use trunks of small diameters, because of the penury of those with a large one, he does not make any deeper structural consideration and tacitly, in some way, he accepts the ideas of Pol Abraham.

The question of the horizontal force, however, is important in another situation: that of the wind action. Placed above so high constructions, the Gothic builders could not ignore that  the wind acting on the roof of a cathedral acted upon the stone structure below with a considerable thrust, impossible to be counterbalanced by friction (studies about the wind on Notre-Dame can be found in \cite{mark84}, with a historical perspective, or in \cite{NDwind}). They needed a safe device to transmit the horizontal thrust of the wind to the upper part of the clerestory. Such a device is composed by the {\it jambes} of the {\it fermes principales} with their {\it aisseliers} and {\it chapeau}, Fig. \ref{fig:8}; in \cite{chevrier}  this set of struts  is indicated as the {\it console}. Contrarily to what normally thought, this device is not introduced to  improve the vertical support of the main frames, the width of the guttering wall is sufficient for that, nor for relieve the bending of the {\it entrait}, as suggested in \cite{fromont}. In fact, to decrease the bending of the {\it entrait} it should be easier and less expensive to use two more {\it poteaux}, i.e. vertical ties. In addition, looking at the {\it fermes} of the nave, we can see that the {\it console} should support the {\it entrait} just where there is already a vertical tie, which is redundant. For the choir's {\it fermes}, the span of the {\it  entrait} is divided into three parts, which is largely sufficient to help the {\it entrait} in bending, as confirmed by the numerical simulations presented in Sect. \ref{sec:stresses}. 
Indeed, the true role of the {\it console} system is to transmit the wind force to the lower part of the guttering wall. Through the inclined {\it aisseliers}, the horizontal thrust flows as an inclined force in the lower part of the guttering wall, through the stone corbels close to the vault level, so improving considerably the strength of the clerestory structure to the action of the wind. This  is corroborated by the presence of a {\it clavette}, i.e. a strong shear key, connecting the {\it entrait} and the {\it chapeau} of the {\it console}, Fig.  \ref{fig:13} a), whose role is to transfer the wind thrust from the {\it entrait } to the {\it console}, while it is completely useless for transmitting vertical forces. In addition, the {\it consoles} transfer a part of the vertical load, also for the only own weight of the structure, to the lower part of the guttering wall. This is not ideal, because, in order to improve the safety of this last, the better is to put the entire vertical load on top of it. It is hence likely that the only reason that pushed the carpenters of the cathedral to introduce the {\it consoles} was to dispose of a good device to safely transmit the  wind action on the roof to the stone structure below. 
We will see in Sect. \ref{sec:structanalysis} that the numerical simulations confirm this fact.

\begin{figure}
\begin{center}
\includegraphics[width=.7\textwidth]{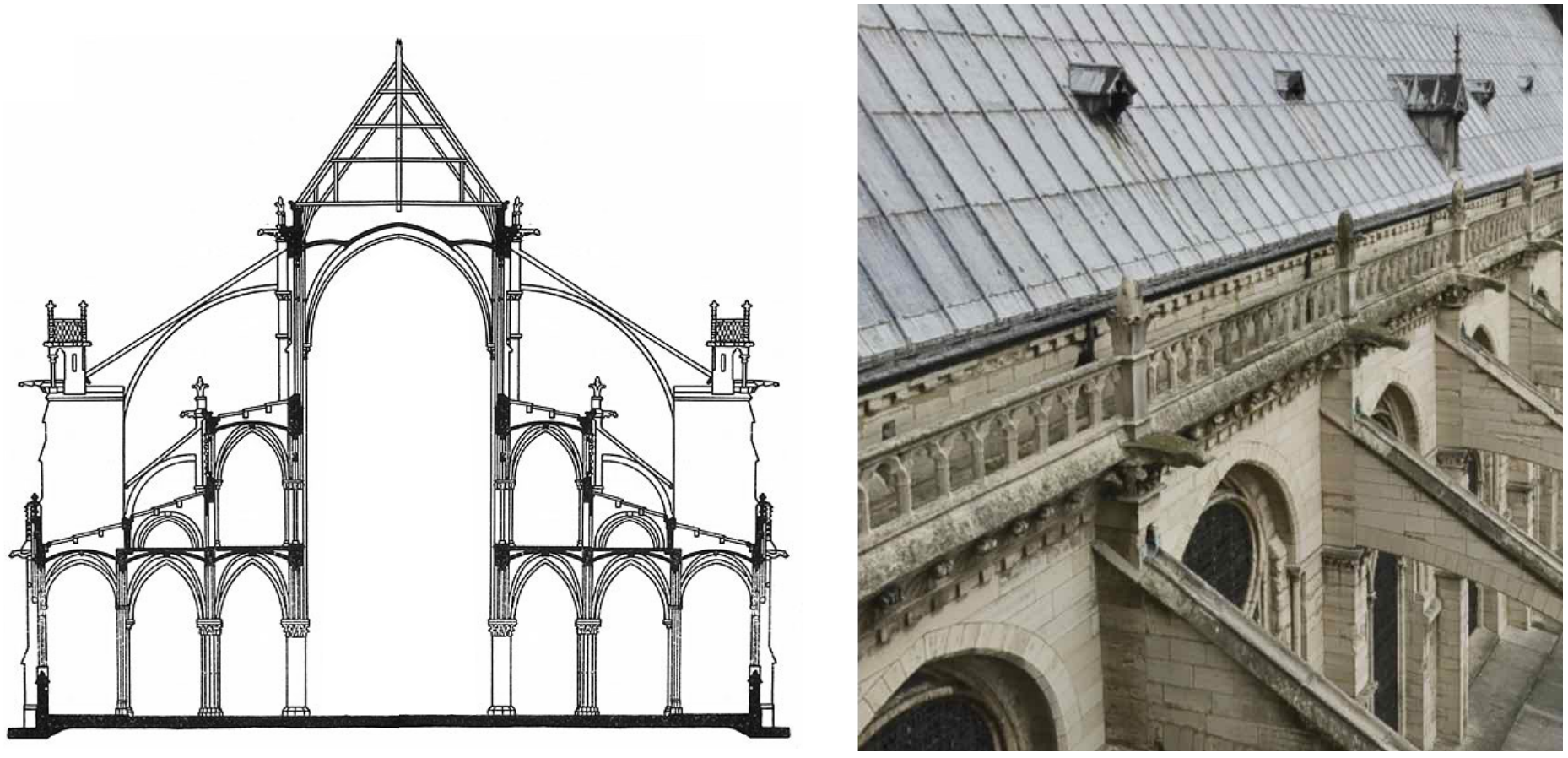}
\caption{The vertical gap between the top of the flying-buttresses and  that of the guttering wall.}
\label{fig:16}
\end{center}
\end{figure}

The \cha\ of a Gothic cathedral was, normally, built before the construction of the high vault, see e.g. \cite{bechmann}. During this phase, the \cha\ had also another role: to counterbalance  the flying-buttresses thrust, ensuring the connection between the two sides of the clerestory before the construction of the vault. The device constituted by the {\it consoles} well assumed this structural role:  the {\it entraits}, equipped with the two {\it consoles}, were able to balance and resist the inward thrust applied to the two opposite clerestory walls by the flying-buttresses on the North and South side of the cathedral. In particular, in Notre-Dame, after the raise of the guttering wall, the top of the flying-buttresses was too far below the \cha\ to assume such a kind of thrust balance without a device, the {\it consoles}, acting down below the \cha. The vertical gap between the \cha\ and the top of the flying-buttresses can be appreciated in Fig. \ref{fig:16}.

The overall behavior of the \cha, conferred by the set of {\it chevrons maîtres}, {\it fermettes} and {\it sablières}, was not sufficient to ensure the equilibrium of the structure. Another element was of the uttermost importance: the brace system ({\it contreventement}, in French). This was composed by longitudinal ties, here denoted, according to \cite{chevrier}, the {\it liernes}, and their braces, the {\it aisseliers}, Figs. \ref{fig:11} and \ref{fig:12}. In the \com\ of the choir of Notre-Dame, there are three sets of {\it liernes} and {\it aisseliers}, in the nave they are four between FN2 and FN5, five in the rest of the \cha\, where a device is added on the top of the charpente and the {\it lierne} coincides with the {\it panne faîtière} already cited; according to \cite{fromont}, this could be a posterior addition, but this point of view is not confirmed. Another possibility, as already said, is an unrepaired rupture. 
In the most part of texts, like in \cite{fromont}, this part of the \cha\ is considered to be used with the task of ensuring the longitudinal equilibrium of the structure. This is likely, especially during the construction phases. However, two considerations must be done. The first one, the bracing role of the longitudinal {\it liernes} and {\it aisseliers} is bounded not only to withstand horizontal forces, like wind or earthquakes, which is the true structural reason for a bracing system, but more importantly to ensure a global stability of the structure against rigid mouvements, namely rotations, in the longitudinal direction. The second one is correctly identified in \cite{epaud1}: the system formed by the {\it liernes} and {\it aisseliers} was not merely, and certainly not mainly, a bracing system, a {\it contreventement} in the proper, structural sens of the term. It was principally a system ideated by the Gothic carpenters to relieve the bending of the {\it faux entraits} of the {\it fermettes} and to transfer an important  part of the vertical load of these ones onto the {\it chevrons maîtres}. Numerical simulations confirm this fact, see Sect. \ref{sec:reactions}. That is why the common idea that the system of the  \cha$s$ with {\it chevrons formant fermes} distributed rather uniformly the vertical load on the top of the guttering wall, see e.g. \cite{viollet, enlart,bechmann} and rather surprisingly also \cite{epaud1}, just after having detected the true role of the longitudinal ties and braces, is wrong. We will see that the structural analysis of the \cha\ clearly shows that the most part of the vertical load of the \cha\ is reported onto the main frames, while the vertical load transferred to the wall directly by the {\it fermettes} is much lower. This confirms the effectiveness of the longitudinal system of {\it liernes} and {\it aisseliers} in transferring to the {\it chevrons maîtres} the vertical loads. In addition, such a structural organization, which confers to the \cha\ a complex three-dimensional static functioning, helps also in decreasing the bending in the {\it chevrons} of the {\it fermettes} and in  the {\it sablières}, because also the horizontal outward thrust at the base of the {\it chevrons} is diminished by such a structural organization. It is interesting to notice that thanks to the use of the longitudinal system of {\it liernes} and {\it aisseliers}, the carpenters did not need to use vertical ties in the {\it fermettes} to sustain the {\it faux entraits}. Finally, with a unique system, they ensured the {\it contreventement}, a support for the {\it faux entraits} of the {\it fermettes}, the transfer of a part of the vertical loads to the main frames and the decrease of the bending in the {\it chevrons} and in the {\it sablières}. 

The  analysis and interpretation of the structural functioning of the \cha\ presented in this Section is, for the while, merely qualitative, inspired by the criticism of the existing literature on the matter and by an observation of the \cha's structure   as a whole, in relation also to the rest of the structure of the cathedral and to the historical phases of its construction. Nevertheless, these considerations must be supported by a quantitative, structural study of the \cha, which is done in the next Section. 


\section{Structural analysis of the \textit{charpente}}
\label{sec:structanalysis}
\subsection{The mechanical models}
\label{sec:mechmod}
\begin{figure}
\begin{center}
\includegraphics[width=\textwidth]{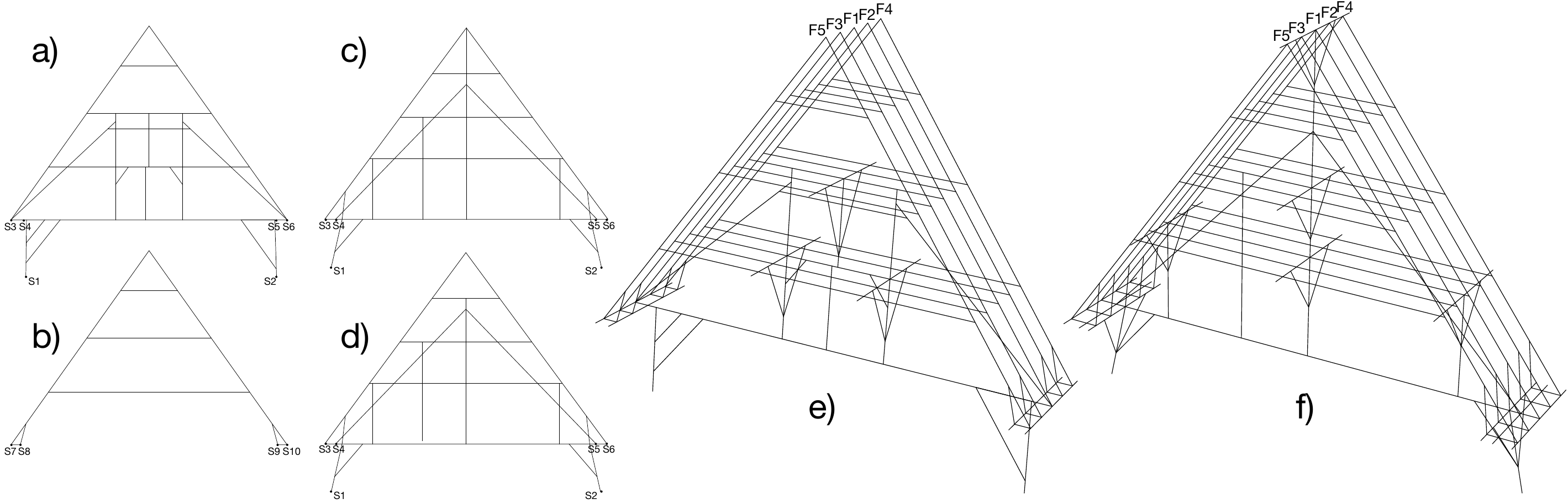}
\caption{Schemes of the \com's structure: a) {\it ferme principale} of the choir; b) a {\it fermette}; c) {\it fermes principales} FN1-FN2 and FN5-FN11 of the nave; d) {\it fermes principales} FN3-FN4  of the nave; e) the structural unit of the choir's \cha; f) the structural unit of the nave's \cha. }
\label{fig:17}
\end{center}
\end{figure}

The \cha$s$ of the choir and of the nave are studied here. Each one is modeled as an assembly of elastic rods, pinned at the ends and at each intermediate intersection with another beam: the  joints between wooden beams of the types used in the medieval structures, like tenon and mortice or {\it mi bois}, are not able to transmit bending. Based upon the data in \cite{chevrier, fromont,laserscan}, the structural unit of the each one of the two  \cha$s$ is transformed into a three-dimensional structural scheme, shown in Fig. \ref{fig:17}, where the points corresponding to the supports are also indicated with labels S1 to S10. All of these points are modeled as frictionless unilateral boundary conditions, where a purely contact reaction can be exerted (corresponding to the physical condition of a contact between two superposed bodies). In particular, at each one of the points from S3 to S10, only upward vertical forces can be exerted by  the support. In S1 and S2, two unilateral reactions are exerted, an inward horizontal one and a vertical upward one. It is important to notice that as a consequence of the deformations produced by the loads, some of the support points can detach from their footing: the boundary conditions of the \cha\ can change according to the loading condition. Typically, this happens when wind loads are added to the own weight. This is another reason for not including friction in the mechanical model.

The structural units of the two \cha$s$ are then transformed into two Finite Element (FE) models. Each element is modeled as an Euler-Bernoulli rod, which is justified by the slenderness of the beams, ties and struts, \cite{sokolnikoff,pmmariano}. The FE model has been established with the aid of a classical FE code, particularly suited for modeling beam structures, SAP. 
The boundary conditions imposed to the points at the ends of the {\it liernes } and {\it sablières} specify a continuity of displacements and rotations with the corresponding elements of the continuous structural units. In this way, the simulation done on a singular structural unit represents  the global structural response of the \cha\ for each one of its parts, in the assumption of uniform loading all over the \cha\, which is the case for the own weight and,  at least to a first approximation, usually accepted in civil engineering for the wind action. 

In the case of the nave, two different calculations are performed: one for the structural unit  in Fig: \ref{fig:17} f), valid for FN1 to FN2 and for FN5 to FN11. The other one, for the structural units of FN3 and FN4, for which the model is slightly modified to account for  some peculiarities, described above: the {\it suspente} is not linked to the {\it entrait}, the {\it poinçon} is interrupted at the level of the third {\it faux entrait} and  the tie connecting together the {\it consoles} is not linked to the {\it jambettes}, Fig. \ref{fig:17} d).

\subsection{ Material characteristics}
As already mentioned, the \cha\ was made with the wood of oak trees. For such a material, when dry, the density can be evaluated at $\sim$710 kg/m$^3$.
Though transversely isotropic, \cite{vannucci}, in this model the wood has been considered as isotropic. In fact, the beams are essentially solicited by bending and axial force, the relevant mechanical characteristic is hence the Young's modulus $E$ in the direction of the wood fibres, here taken equal to 12500 MPa.  The Poisson's coefficient has been put equal to 0.25.

\subsection{Dimensions of the beams}
The dimensions of the beams composing the \cha\  are reported in Tab. \ref{tab:1}. For the few lacking data, the dimensions of the sections have been deduced from the observation of some  pictures of the \com. The minimum diameter $d_{min}$ of a trunk to obtain the corresponding cross section is also indicated in Tab. \ref{fig:1} for all the parts of the \cha. As observed in \cite{epaud1}, the most part of beams, ties and struts can be obtained by trunks with a moderate diameter, less than $\sim30\div35$ cm.
Above the \com, as already mentioned, there was the {\it voligeage}, made of fir wood (density $\sim$ 500 kg/m$^3$); for it, a thickness of 2 cm has been considered. 
Finally,  the catwalk is to be considered too. Wider in the nave than in the choir, its mass can be evaluated to $\sim240$ kg for each structural unit in both the cases. Finally, we get the results presented in Tab. \ref{tab:2} concerning the global volumes and masses of wood for each structural unit.

\begin{table}[htp]
\caption{Dimensions of the wood beams, as deduced from \cite{fromont}; for each section, $b$ is the width,  $h$ its height, $A$ the area, $J_1$ and $J_2$ the moments of inertia, $J_0$ the polar moment of inertia and $\mu$ the linear density of mass.}
\begin{small}
\begin{center}
\begin{tabular}{lrrrrrrrr}
\hline
\multicolumn{9}{c}{Choir's \cha}\\
\hline
	&$ b$ & $ h$   & 	$d_{min}$ &	$A$ &	$J_1$  & 	$J_2$  & 	$J_0$ & 	$\mu$ \\
	& [cm]   &      [cm]       &    [cm]         &            [cm$^2$]    &  [cm$^4$] & [cm$^4$]  &  [cm$^4$]  &  [kg/m]\\
	\hline
{\it Entrait} &	30	& 35	& 46.1	& 1050	& 107188 &	78750	&185938	& 74.550\\
1st {\it faux entrait, liernes}	&13&	27	&30.0&	351&	21323&	4943	&26267&	24.921\\
2nd {\it faux entrait} &	17	&19.5	& 25.9	& 332	& 10504 &	7984	 & 18488 &	23.537\\
3rd {\it faux entrait} & 	15 & 	23	& 27.5 &	345	& 15209	& 6469 &	21678 &	24.495\\
4th {\it faux entrait} &	15 &	19 &	24.2	& 285 &	8574 & 	5344	 & 13918	& 20.235\\
{\it Arbalétriers} &	18	& 19	& 26.2	& 342 &	10289 &	9234	 & 19523 &	24.282\\
{\it Faux arbalétriers} &	28	& 17	& 32.8 &	476 &	11464 &	31099 &	42562 &	33.796\\
{\it Poteaux} &	19 &	15 &	24.2	 & 285	& 5344 &	8574	& 13918	& 20.235\\
{\it Poteau central haut} &	14	&14	&19.8&	196	&3201&	3201&	6403	&13.916\\
{\it Aisseliers faux entrait} &	14&	17&	22.0&	238&	5732	&3887	&9619&	16.898\\
{\it Jambe gauche, jambettes}&	18	&23	&29.2	&414&	18251&	11178&	29429&	29.394\\
{\it Aiss.   j. gauche} and  {\it  liernes}	&14	&18	&22.8&	252&	6804	&4116&	10920&	17.892\\
{\it Jambe droite}&	30&	19&	35.5	&570	&17148&	42750	&59898	&40.470\\
{\it Aisselier jambe droite}&	30	&18	&35.0&	540	&14580&	40500&	55080&	38.340\\
{\it Blochet }&	15	&15	&21.2&	225	&4219&	4219	&8438	&15.975\\
{\it Sablières}&	19&	14&	23.6&	266	&4345&	8002	&12347&	18.886\\
\hline
\multicolumn{9}{c}{Nave's \cha}\\
\hline
{\it Entrait}	&26&	29&	38.9&	754&	52843&	42475&	95318&	53.534\\
{\it Faux entraits}&	17&	24&	29.4&	408&	19584&	9826&	29410&	28.968\\
{\it Arbalétriers}	&16&	25.5&	30.1&	408&	22109&	8704&	30813&	28.968\\
{\it Faux arbalétriers}&	17&	19&	25.5&	323&	9717	&7779&	17496&	22.933\\
{\it Poinçon}&	23.5&	18.5&	29.9&	435&	12399&	20008&	32407&	30.867\\
{\it Suspente}&	12&	12&	17.0&	288&	3456&	3456&	6912&	20.448\\
{\it Poteaux}&	17&	20&	26.2&	340&	11333&	8188	&19522&	24.140\\
{\it Jambettes}&	15&	16&	21.9&	240&	5120	&4500&	9620&	17.040\\
{\it Liernes}&	15&	18&	23.4&	270&	7290&	5063&	12353&	19.170\\
{\it Aisseliers} and {\it blochets}&	15&	15&	21.2&	225&	4219&	4219&	8438&	15.975\\
{\it Jambe de force}&	20&	15&	25.0&	300&	5625&	10000&	15625&	21.300\\
{\it Chevrons secondaires}	&17&	24&	29.4&	408&	19584&	9826	&29410&	28.968\\
{\it Sablières}&	22&	14&	26.1&	308&	5031&	12423&	17453&	21.868\\
\hline
\end{tabular}
\end{center}
\end{small}
\label{tab:1}
\end{table}

\begin{table}[htp]
\caption{Global quantities of wood for the structural units of choir and nave.}
\begin{small}
\begin{center}
\begin{tabular}{lrrrrr}
\hline
&\multicolumn{2}{c}{Choir}&&\multicolumn{2}{c}{Nave}\\
\cmidrule{2-3}\cmidrule{5-6}
&Mass &Volume& &Mass &Volume \\
&[kg]&[m$^3$]&&[kg]&[m$^3$]\\
\hline
{\it Ferme principale}& 3168& 4.46&&2920 &4.11 \\
{\it Fermette}&1050 &1.48 &&1220 &1.72 \\
{\it Contreventement } and {\it sablières}&856 &1.21 &&1190 &1.68 \\
Total for the structure&5074 & 7.15&&5330 &7.51 \\
{\it Voligeage}&920&1.84  &&766&1.53\\
{\it Passerelle}&240&0.34 && 240&0.34\\
Total for the structural unit & 6234&9.33 &&6336 &9.38 \\
Total per unit  length&1520&2.27&&1810&2.68\\
\hline
\end{tabular}
\end{center}
\end{small}
\label{tab:2}
\end{table}

It can be noticed that the quantity of wood is almost the same for the two structural units, but  the total mass per unit length measured along the cathedral axis is $\sim$19\% greater for the nave's \cha. It is often affirmed that the \cha\ of the nave is more perfected than that of the choir, but in consideration of these data, what can be said is that the change of the structural scheme was not dictated by economical considerations: the reason is another one. The main frame of the nave has, actually, a better mechanical behavior than that of the choir. Roughly speaking, it needs a less quantity of matter to obtain the same stiffness and, in fact, the nave's {\it chevrons maître} is lighter than the choir's one. But on the whole, the nave's \cha\ is   heavier than that of the choir. This is  due to the greater weight of the bracing system and of the {\it fermettes} of the nave. 
If the differences between the two \cha$s$ are attentively considered, it can be seen that all the structural action  of the nave's designer is oriented to increase the stiffness, and so the stability, of the \cha: all the structural changes made with respect to the choir, i.e. the change of static scheme for the main frame, the reinforcement of the bracing system and the greater sections used for the {\it fermettes}, go in the direction of  a stiffer structure. 
It is not, in fact, the strength that  is substantially improved, because the level of stresses in the two \cha$s$ is similar, Sect. \ref{sec:stresses}, but the  stiffness of the structure, as confirmed by a comparative modal analysis,  Sect. \ref{sec:dynamic}. 
While a stress analysis was with no doubts out of the means of the builders of the Middles Ages (the concept of stress is introduced by Cauchy in the XIXth century, \cite{benvenuto}), an embryonic perception of the stability, and hence of the stiffness, of a structure can have been in the abilities of the Gothic carpenters. It can be acquired by experience, especially during the construction phases  and, thanks to this experience and ability, some particularly perceptive carpenters can have improved the technique, like in the case of Notre-Dame.

\subsection{Loading conditions}
\begin{figure}
\begin{center}
\includegraphics[width=.6\textwidth]{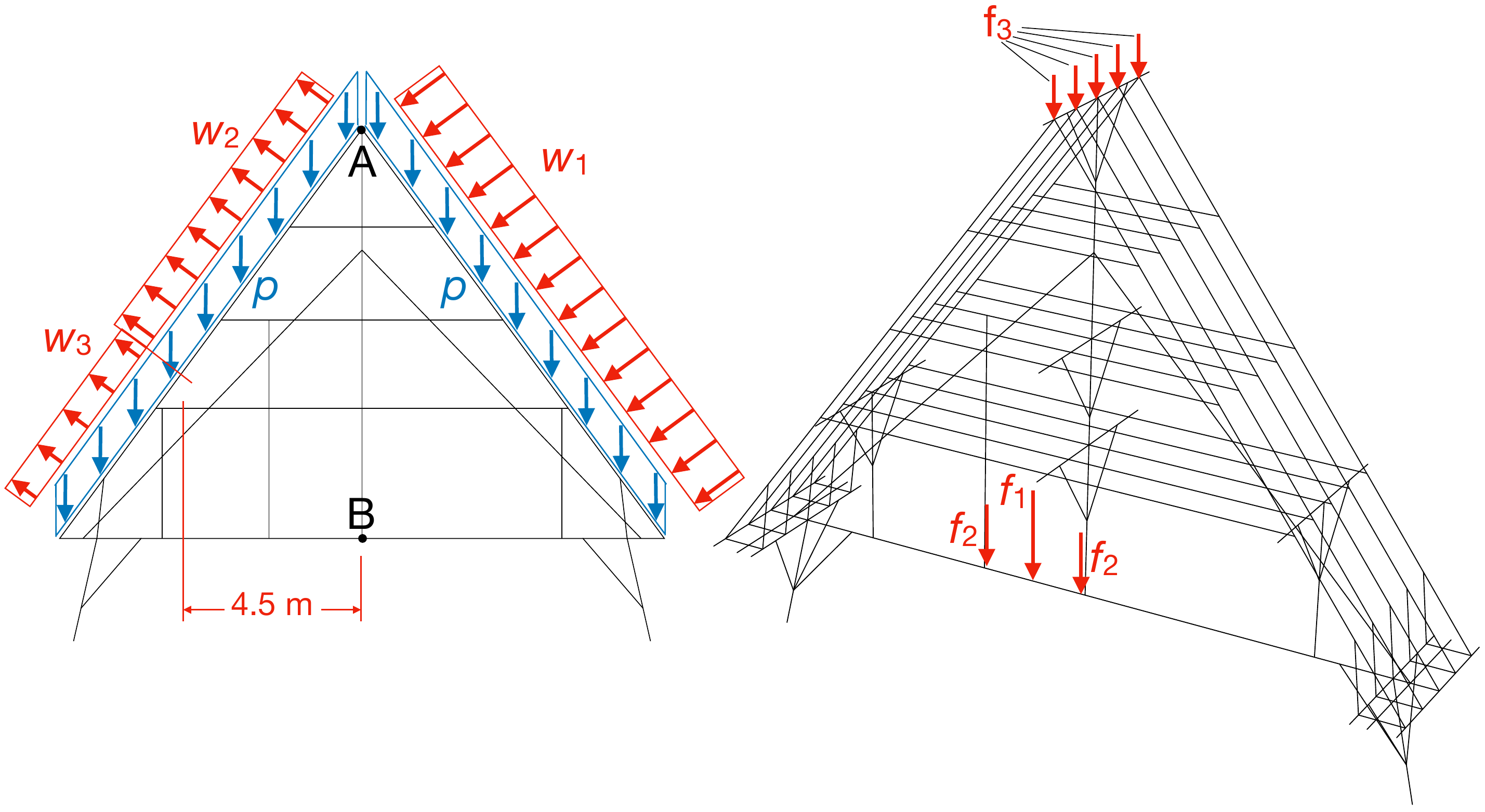}
\caption{Scheme of the actions on the \cha; $p$: lead tiles and {\it voligeage} load; $w_1,w_2,w_3$: wind load; $f_1,f_2$: load of the catwalk; $f_3$: load of the frieze. }
\label{fig:19}
\end{center}
\end{figure}

Two loading conditions have been considered: own weight and own weight plus the  wind. For what concerns the own weight, two different cases have been analyzed:  the original state (OS), i.e. the one before the changes occurred in 1726, and the final state (FS), that before the fire of April 15th, 2019. The differences between these two cases are due to the lead covering: tiles with a thickness of 5 mm for the OS, of 2.82 mm for the FS and the presence of the frieze, Fig. \ref{fig:7}. The OS is analyzed to evaluate the structural situation of the original \cha\, as designed by the Gothic carpenters, while the FS is considered to have an assessment of the state of the \cha\ before its destruction. 
For what concerns the wind loads, the objective is just to evaluate what were the consequences on the \cha\ of an horizontal load. This analysis is a rather classical one: the wind action is modeled as a static load, orthogonal to the impinged surface, and distributed on the windward (overpressure) and leeward (suction) sides of the roof. The value of the wind load is calculated and distributed according to the  European norm EUROCODE 1, \cite{eurocode}. The objective, however, is not to evaluate whether or not the \cha\ was safe according to such a  standard, a meaningless objective for an ancient and now destroyed structure, but just to have a value of the wind action that is commonly accepted in the analysis of civil structures. 
The actions applied to the \cha\ are sketched in Fig. \ref{fig:19} and the values of the loads shown in  Tab. \ref{tab:3}.

\begin{table}[htp]
\caption{Loads on the \cha; $p_0$ is the linear load of the {\it voligeage}, $p_1$ that of the lead tiles and $p$ their sum; for the other symbols, refer to Fig. \ref{fig:19}.}
\begin{small}
\begin{center}
\begin{tabular}{lrrrrrrrrrrrrrr}
\hline
&&&&&&&\multicolumn{3}{c}{Original state}&&\multicolumn{3}{c}{Final state}\\
\cmidrule{8-10}\cmidrule{12-14}
&$w_1$&$w_2$&$w_3$&$p_0$&$f_1$&$f_2$&$f_3$&$p_1$&$p$&&$f_3$&$p_1$&$p$\\
\cmidrule{2-4}\cmidrule{6-7}\cmidrule{9-10}\cmidrule{13-14}
&\multicolumn{3}{c}{[N/m]}&[N/m]&\multicolumn{2}{c}{[N]}&[N]&\multicolumn{2}{c}{[N/m]}&&[N]&\multicolumn{2}{c}{[N/m]}\\
\hline
Choir&480.1&351.4&269.4&80.4&947&700& - &479&559.4&&467&283&363.4\\
Nave&410&300&230&68.7&1037&668&-&408.8&477.5&&398.7&241.6&310.3\\
\hline
\end{tabular}
\end{center}
\end{small}
\label{tab:3}
\end{table}

\subsection{Validation of the FE model}
\begin{figure}[htp]
\begin{center}
\includegraphics[width=\textwidth]{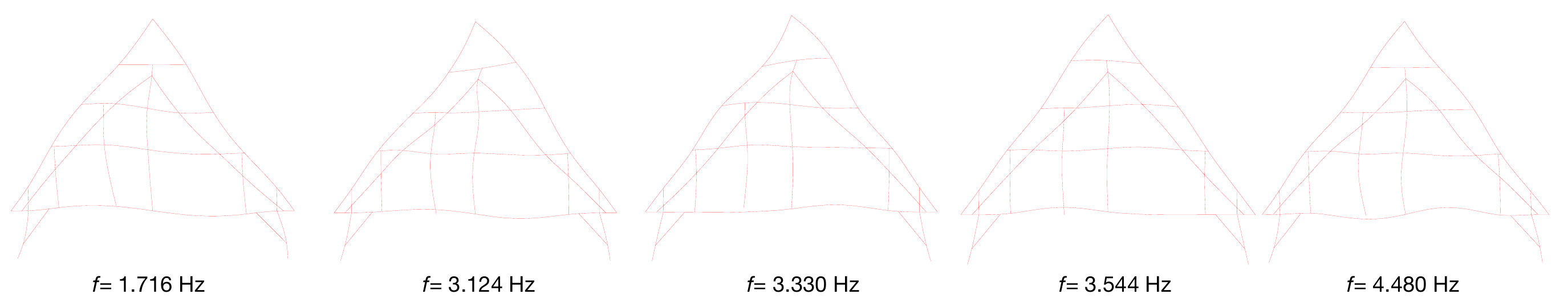}
\caption{From left to right: the first five modal forms of FN3, CAST3M simulation. }
\label{fig:20}
\end{center}
\end{figure}

The simulations of the structural response of the \cha\ have been done in the framework of linear elasticity, which is justified by the context: no nonlinear effects due to the response of the material are to be expected on such a type of structure (the wood is viscoelastic, but after a so long period, all the viscid deformation was undoubtedly ended) as well as non linearities of the geometrical type, like buckling or snap-through. An elastic linear analysis is hence appropriate in this context; the FE method is well established for such a type of study and normally there is no need for convergence analysis. Anyway, in order to evaluate the quality of the FE model used for the structural analysis of the \cha, a control of the quality of the simulation has been done on a simplified model. This is a model of the FN3 of the nave, that has been analyzed with SAP, using a coarse model (122 Degrees of Freedom, DoF), and with the  more sophisticated code CAST3M, \cite{castem}, using a refined mesh (70860 DoF), in both the cases considering exclusively the own weight of the \cha. The comparison between the results of the two calculations, done on the values of the reactions, on the displacement of points A and B in Fig. \ref{fig:19} and on the value of the frequencies of the first five global normal modes, is given in Tab. \ref{tab:4}.  As apparent, the results obtained with the coarse model are extremely close to those given by the refined model; the only remarquable difference concerns the  frequencies of modes 4 and 5. This is quite normal: to catch with precision high frequency modes the fineness of the mesh is important. However, what matters in this problem is the fundamental frequency, mode 1, which is well predicted by the coarse model. Hence, the use of a coarse FE model for this problem is not prejudicial of the quality of the results.
For the sake of completeness, the modal forms of FN3 are shown in Fig. \ref{fig:20}.
Finally, for the whole structural unit of the nave's \cha\ a model with 852 DoF has been used, while for the choir's one the  model had 2347 DoF.
\begin{table}[htp]
\caption{Comparison of the results  for the SAP and CAST3M models of FN3.}
\begin{small}
\begin{center}
\begin{tabular}{lrrrrr}
&\multicolumn{2}{c}{SAP}&&\multicolumn{2}{c}{CAST3M}\\
\cmidrule{2-3}\cmidrule{5-6}
&\multicolumn{5}{c}{Reaction forces [N]}\\
\hline
Node&$R_x$&$R_y$&&$R_x$&$R_y$\\
\hline
S1&1011.6&2556.8&&1011.8&2557.3\\
S2&-773.6&2068&&-773.5&2067.7\\
S3&3700.6&4227.7&&3700.6&4227.4\\
S4&4801&7629.6&&4801&7630\\
S5&-5203.3&7795&&-5203.6&7795\\
S6&-3536.3&4198&&-3536.3&4198.3\\
\hline\smallskip\\
&\multicolumn{5}{c}{Nodal displacements [mm]}\\
\hline
Node&$\delta_x$&$\delta_y$&&$\delta_x$&$\delta_y$\\
\hline
A&$-4.63\times10^{-3}$&$-8.02\times10^{-2}$&&$-4.63\times10^{-3}$&$-8.04\times10^{-2}$\\
B&$1.10\times10^{-4}$&$-2.47\times10^{-1}$&&$1.13\times10^{-4}$&$-2.47\times10^{-1}$\\
\hline\smallskip\\
Mode&\multicolumn{5}{c}{Frequencies of vibration [Hz]}\\
\hline
1&\multicolumn{2}{c}{1.749}&&\multicolumn{2}{c}{1.716}\\
2&\multicolumn{2}{c}{3.138}&&\multicolumn{2}{c}{3.124}\\
3&\multicolumn{2}{c}{3.462}&&\multicolumn{2}{c}{3.330}\\
4&\multicolumn{2}{c}{4.945}&&\multicolumn{2}{c}{3.544}\\
5&\multicolumn{2}{c}{6.181}&&\multicolumn{2}{c}{4.480}\\
\hline
\end{tabular}
\end{center}
\end{small}
\label{tab:4}
\end{table}

\subsection{Results of the numerical simulations}
\label{sec:results}

As already mentioned, for each \cha, i.e for the choir and for the two cases of the nave (ordinary frames and frames FN3-FN4), four different calculations have been done: with only the own weight for the load (denoted hereafter OW) or with also the wind (denoted by OW+W), both of them for the original state, i.e. the state of the \cha\ before the restorations of 1726, hence with lead tiles of 5 mm of thickness, and the final state, the one after the restoration of Lassus and Viollet-le-Duc, with lead tiles of 2.82 mm thick and the frieze in Fig. \ref{fig:7}. The results of the numerical calculations made on the FE models described above are reported below. The distribution of the reaction forces, of the displacements and of the stresses are presented successively.

\subsubsection{Reaction forces}
\label{sec:reactions}
\begin{figure}[h]
\begin{center}
\includegraphics[width=.85\textwidth]{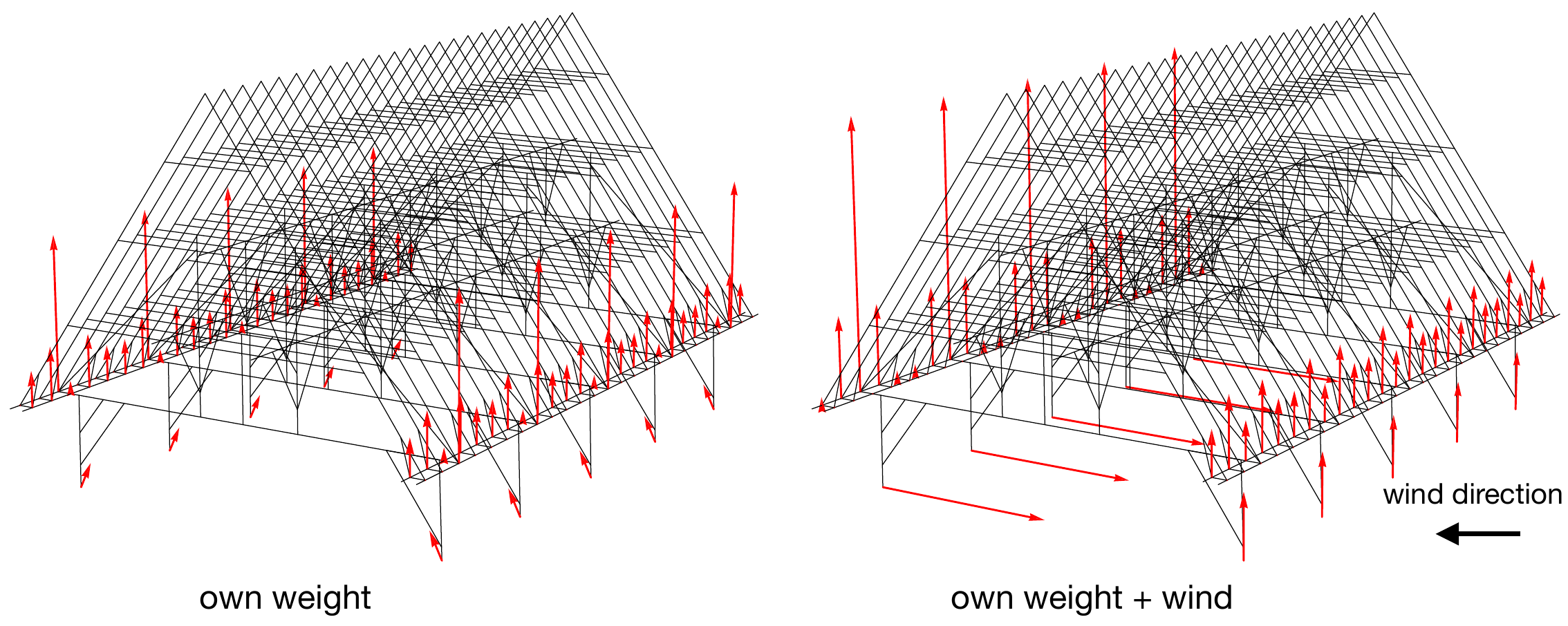}
\includegraphics[width=\textwidth]{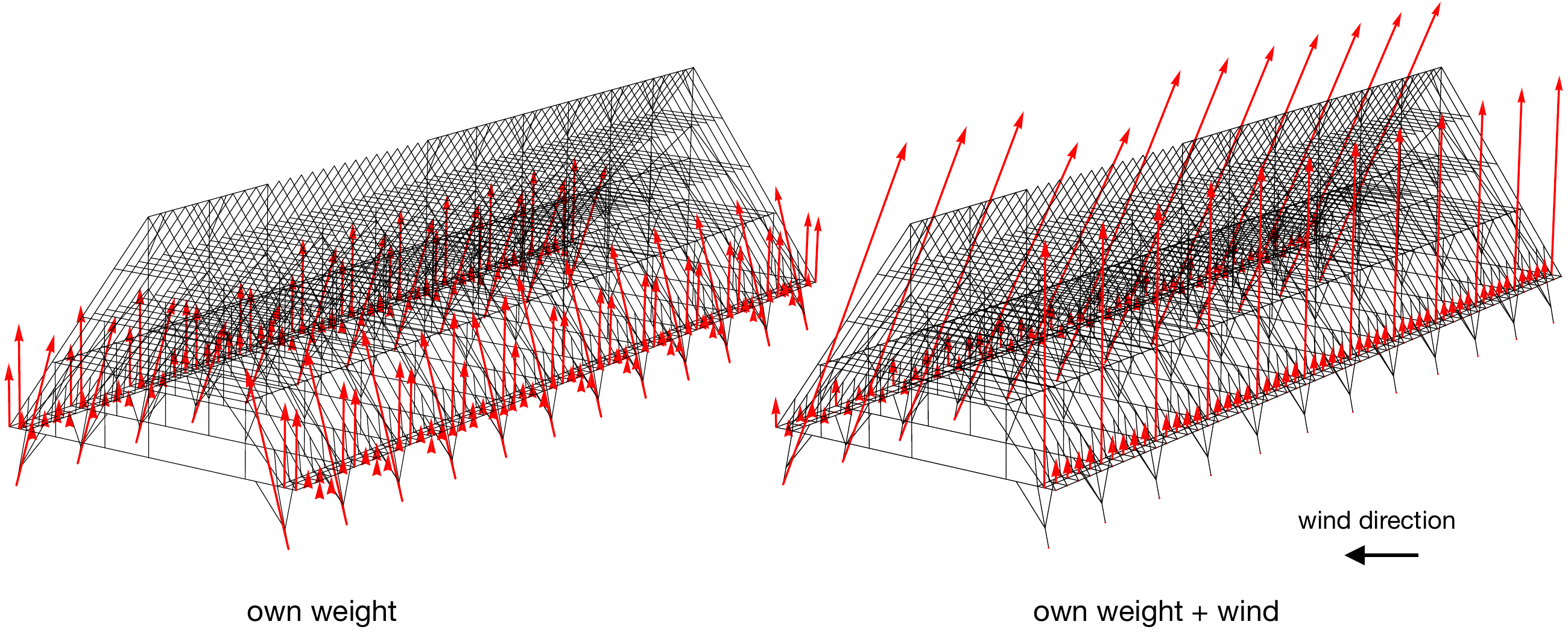}
\caption{Distribution of the reaction forces in the original state; top, the  choir's \cha, bottom the nave's one. }
\label{fig:21}
\end{center}
\end{figure}

The distribution of the reaction forces for each support point, S1 to S6 for the main frame F1, S7 to S10 for the secondary frames F2, F3 and F4, F5, cf. Fig. \ref{fig:17}, is detailed in Tab. \ref{tab:5} for all the \cha$s$ and loading conditions ($R_x$: horizontal reaction; $R_y$: vertical reaction).
The distribution of the reactions is almost symmetric between the South and North sides, though not exactly, due to the small asymmetries in the main frames structure. It is interesting to notice that not all the supports are active for the vertical load (condition OW): for the choir, the supports S7 and S10 of the {\it fermettes} are inactive, as well as S9 for F2 and F3 of the nave's \cha\, exception made for the part between FN3 to FN4. We can also observe that the vertical reactions in S1 and S2, i.e. at the level of the corbels supporting the {\it jambes}, is  far less than that absorbed by nodes S3 to S6, on the guttering wall's top, for the choir's \cha, while it is higher in the nave, due to the different arrangement of the structure. This confirms that the system of wooden {\it consoles} is not determining in supporting the vertical loads, as already commented. 
For the loading condition OW+W, the situation is different: the presence of the {\it jambes} strongly affects  the distribution of the reactions. The set of supporting nodes changes: in the choir, nodes S3 on the North side (the wind is simulated to blow from the South, where the cathedral is more exposed to the wind) are heavily charged, while nodes S4, S5, S8 and  S10 are inactive. Nodes S6, S7 and S9 are also active, but with a reaction force far below that of S3. The node S1 is charged only horizontally, to entirely absorb the wind thrust, while S2 is charged in the vertical direction, as an effect of the roof's slope, see below.
In the nave, due to the different scheme, S2, S4, S6, S8 and S10 are inactive, while S1, S3, S5, S7 and S9 active. It is interesting to notice the high reaction at node S5, consequence of the structural scheme. 
The distribution of the reaction forces is depicted in Fig. \ref{fig:21}, for the original state. Better than the data in Tab. \ref{tab:5}, this two figures let see how the distribution of the reactions on the top of the guttering walls was far from being uniform. The main part of the load is transmitted by the {\it chevrons maîtres}, that play a fundamental role in the structural functioning of the \com. 

\begin{sidewaystable}[htp]
\caption{Reaction forces for the choir and nave structural units.}
\begin{small}
\begin{center}
\begin{tabular}{lrrrrrrrrrrrrrrrr}
\hline
&&&\multicolumn{4}{c}{Choir}&&\multicolumn{4}{c}{Nave's FN1 to FN2 and FN5 to FN11}&&\multicolumn{4}{c}{Nave's FN3 to FN4}\\
\cmidrule{4-7}\cmidrule{9-12}\cmidrule{14-17}
&&&\multicolumn{2}{c}{OW}	&\multicolumn{2}{c}{OW+W}&&\multicolumn{2}{c}{OW}&\multicolumn{2}{c}{OW+W}&&	\multicolumn{2}{c}{OW}&\multicolumn{2}{c}{OW+W}\\
\cmidrule{4-7}\cmidrule{9-12}\cmidrule{14-17}
Frame&Node&	Force  [N]	&OS&FS&OS&	FS&&OS&FS&OS&FS&&OS&FS&OS&FS\\
\hline

\multirow{12}*{F1}&\multirow{2}*{S1}&$R_x$&	2472&	2356&	37020&	37020&&	9383&	8325	&30260	&30260&	&8351&	7388	&30260&	30260\\
&&$R_y$&	5084&	4912&	0&	0&&	30220&	26700&	69940&	67680&&	21970&	19470&	61120&	59910\\
&\multirow{2}*{S2}&$R_x$&	-2472	&-2356	&0&	0&	&-9383&	-8325&	0	&0&&	-8351&	-7388&	0&	0\\
&&$R_y$&	4934	&4716	&12310	&15520&&	31100&	27580&	0	&0&	&27370	&24200	&0&	0\\
&\multirow{2}*{S3}&$R_x$&	0&	0	&0&	0&	&0	&0	&0&	0&	&0&	0	&0	&0\\
&&$R_y$&	29830&	26510&	51980&	48140&&	12090&	10630&	5984	&2125&&	13750	&11830&	6735&	2855\\
&\multirow{2}*{S4}&$R_x$&	0	&0	&0	&0&&	0&	0&	0&	0&&	0	&0	&0&	0\\
&&$R_y$&	2481	&2629&	0	&0&&	19900&	18110&	0&	0&&	13230&	12830&	0&	0\\
&\multirow{2}*{S5}&$R_x$&	0&	0	&0	&0&	&0	&0&	0&	0&&	0&	0&	0&	0\\
&&$R_y$&	2322&	2553&	0&	0&&	15450	&13680&	45600&	39130&&	21060&	19090	&46810&	40300\\
&\multirow{2}*{S6}&$R_x$&	0&	0	&0	&0&	&0	&0&	0	&0&	&0&	0	&0&	0\\
&&$R_y$&	30520&	27170	&7106	&53	&&14780&	13290&	0	&0&	&12270&	10750&	0&	0\\
\hline
\multirow{8}*{F2, F3}&\multirow{2}*{S7}&$R_x$&	0&	0	&0&	0&&	0	&0&	0&	0&	&0	&0&	0	&0\\
&&$R_y$&	0&	0&	15630&	13330&&	2206	&1846&	724	&14&	&0&	0&	1597&	753\\
&\multirow{2}*{S8}&$R_x$&	0&	0	&0	&0	&&0	&0	&0	&0&	&0	&0&	0&	0\\
&&$R_y$&	10230&	8301	&0&	0&	&453&	416&	0&	0&	&5081	&4300&	0	&0\\
&\multirow{2}*{S9}&$R_x$&	0&	0&	0&	0&&	0&	0&	0&	0&&	0&	0&	0&	0\\
&&$R_y$&10190&	8257&	13000&	11050&&	0&	0&	4621&	4196	&&483&	363&	3973&	3565\\
&\multirow{2}*{S10}&$R_x$&	0&	0&	0&	0&&	0&	0&	0&	0&&	0&	0&	0&	0\\
&&$R_y$&	0&	0&	0&	0&&	2873&	2527	&0&	0&&	1651&	1413&	0&	0\\
\hline
\multirow{8}*{F4, F5}&\multirow{2}*{S7}&$	R_x$&	0&	0&	0&	0&&	0&	0&	0&	0&&	0&	0&	0&	0\\
&&$R_y$&	0&	0&	1540&	1014&&	0&	0&	2736&	2565&&	0&	0&	5330	&4790\\
&\multirow{2}*{S8}&$R_x$&	0&	0&	0&	0&&	0&	0&	0&	0&&	0&	0&	0&	0\\
&&$R_y$&	6892	&5583&	0&	0&&	2046&	1742&	0&	0&&	5765&	4868&	0&	0\\
&\multirow{2}*{S9}&$R_x$&	0&	0&	0&	0&&	0&	0&	0&	0&&	0&	0&	0&	0\\
&&$R_y$&	6883	&5573	&8362&	7161&&1587&	1284&	4287&	3732&&	2221&	1834&	3938&	3374\\
&\multirow{2}*{S10}&$R_x$&	0&	0&	0&	0&&	0&	0&	0&	0&&	0&	0&	0&	0\\
&&$R_y$&	0	&0	&0&	0&&	43&	13&	0&	0&&	0&	0&	0&	0\\
\hline
\multirow{7}*{Resultants}&\multicolumn{2}{l}{ $R_x$}&	0&0&37020&37020&&	0&0&30260&30260&&	0&0&30260&30260\\
&\multicolumn{2}{l}{ $R_y$}&143561&123918&148460&128823&&141956&125646&146260&129949&&	140052&123726&144341&128029\\
&\multicolumn{2}{l}{ $R_y$ South side}&71922&	62099&62140&	51995&&70336&	62198&63416&	54986&&69410&61260&62632&	54178\\
&\multicolumn{2}{l}{ $R_y$ North side}&71639&61819&86320&76828&&71620&63448&82844&74963&&70642&62466&81709&73851\\
&\multicolumn{2}{l}{ $R_y$ on walls' top only}&133543&114290&	136150&113303&&80636&	71366&76320&	62269&&90712&	80056&83221&	68119\\
&\multicolumn{2}{l}{ $R_y$ on South wall's top}&	66988&57383&	49830&36475&&	39236&34618&	63416&54986&	&42040&37060&	62632&54178\\
&\multicolumn{2}{l}{ $R_y$ on North wall's top}&	66555&56907&	86320&76828&	&41400&36748&	12904&7283&&48672	&42996&20589&13941\\
\hline
\end{tabular}
\end{center}
\smallskip
The number of the frames and  nodes is indicated in Fig. \ref{fig:17} \\
OW: Own Weight; OW+W: Own Weight+Wind. 
OS: Original State (5 mm lead tiles); FS: Final State (2.82 mm lead tiles and frieze).
\end{small}
\label{tab:5}
\end{sidewaystable}

To remark that the slope of the roof, $\sim55^\circ$, ensures a stabilizing moment of the wind force distribution $w_1$ on the windward side, Fig. \ref{fig:19}, which explains the positive reaction in nodes S2, S6 and S9 of the choir and the high reaction at node S5 of the nave's \cha. Of course, we cannot affirm with certitude that the Gothics were conscious of this fact (the moment of $w_1$ becomes an overturning one at a slope of $60^\circ$), though the concept of moment of a force and its role in the equilibrium of a mechanical system was already roughly  grasped  in the Middle Ages, see \cite{benvenuto}. Nevertheless, the slope of the \com\ has at the same time a stabilizing effect for the wind actions and allows to decrease the bending of the {\it chevrons}, two real structural advantages.
In Tab. \ref{tab:5} the resultant of the reactions is also indicated. This gives also an estimation of the global loads and weights; they are summarized in Tab. \ref{tab:6}.  

\begin{table}[htp]
\caption{Global loads on a structural unit  of  \cha, in [N]. In small: the values per unit length, in [N/m]. $W_x,W_y$: horizontal and vertical components of the total wind force, respectively.}
\begin{small}
\begin{center}
\begin{tabular}{lrrrrrrrrrrr}
\hline
&\multicolumn{2}{c}{ Weight of a SU}&&\multicolumn{5}{c}{Load on wall's top}&&\multicolumn{2}{c}{Total wind force}\\
\cmidrule{2-3}\cmidrule{5-9}\cmidrule{11-12}
& &&&\multicolumn{2}{c}{South wall}&&\multicolumn{2}{c}{North wall}\\
\cmidrule{5-6}\cmidrule{8-9}
&OS&FS&&OS&FS&&OS&FS&&$W_x$&$W_y$\\
\hline
\multirow{2}*{Choir}&143561 &123918 &&66988 &57383 &&66555 &56907 &&37020 &4900\\
			      &\scriptsize{35015} &\scriptsize{ 30224} &&\scriptsize{ 16338} &\scriptsize{ 13996} &&\scriptsize{ 16233} &\scriptsize{ 13880} &&\scriptsize{ 9029} &\scriptsize{ 1195}\\
\hline
{Nave}&141956 &125646 &&39236 &34618 &&41400 &36748 &&30260 &4304\\
 \scriptsize{(FN1-FN2, FN5-FN11)} &\scriptsize{40559} &\scriptsize{35899} &&\scriptsize{11210} &\scriptsize{9891} &&\scriptsize{11829} &\scriptsize{10449} &&\scriptsize{8646} &\scriptsize{1230}\\
 \hline
Nave & 140052 &123726 &&42040 &37060 &&48672 &42996 &&30260 &4304\\
\scriptsize{(FN3 and FN4)} &\scriptsize{40015} &\scriptsize{35350} &&\scriptsize{12011} &\scriptsize{10589} &&\scriptsize{13906} &\scriptsize{12285} &&\scriptsize{8646} &\scriptsize{1230}\\
\hline
\end{tabular}
\end{center}
\end{small}
\label{tab:6}
\end{table}

\subsubsection{Displacement field}
\begin{figure}
\begin{center}
\includegraphics[width=\textwidth]{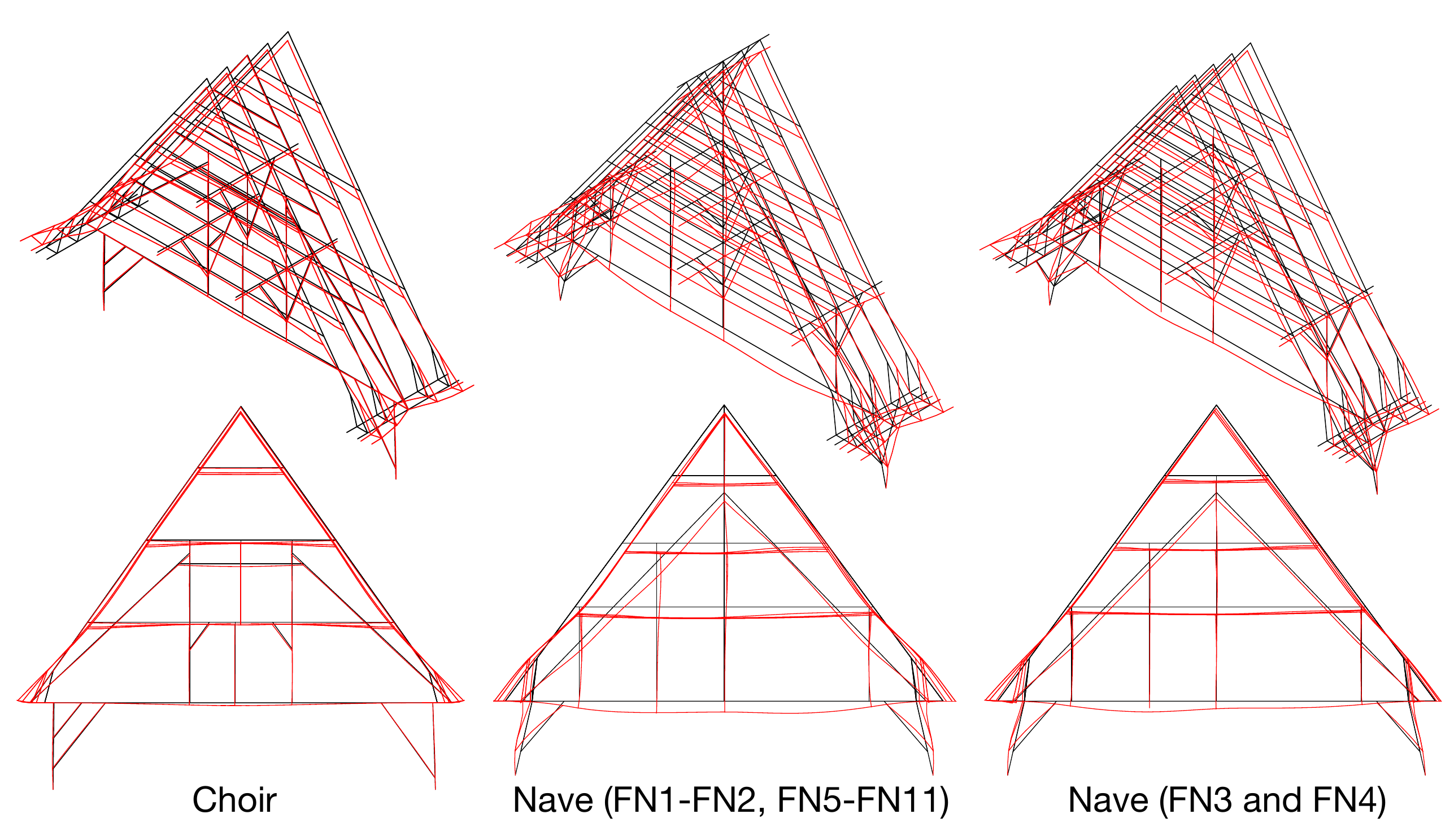}
\caption{Deformation of the structural units of the \cha, in the original state, for the own weight (displacements magnified). }
\label{fig:23}
\end{center}
\end{figure}

\begin{figure}
\begin{center}
\includegraphics[width=\textwidth]{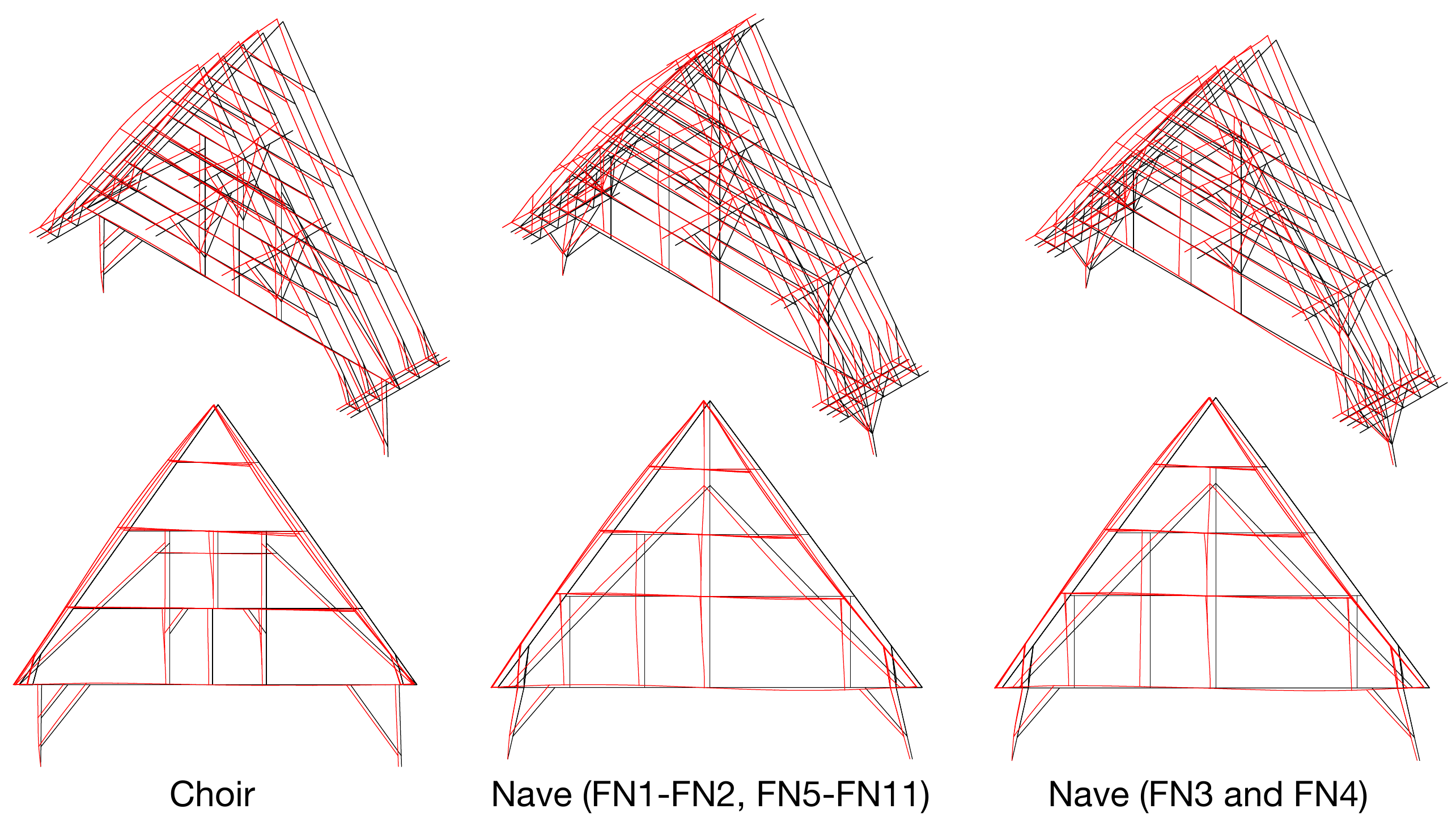}
\caption{Deformation of the structural units of the \cha, in the original state, for the own weight and wind (displacements magnified).}
\label{fig:24}
\end{center}
\end{figure}

The deformation of the structural units of the \com\ is shown in Figs. \ref{fig:23} and \ref{fig:24}, for the case of the original state.
In Tab. \ref{tab:7} the  displacements in the horizontal and vertical direction for the points A and B indicated in Fig. \ref{fig:19} are shown. The entity of the displacements is very small in all the cases.

\begin{table}[htp]
\caption{Displacements of points A and B in Fig. \ref{fig:19} for the orignal state of the \com, [mm].}
\begin{small}
\begin{center}
\begin{tabular}{lrrrrrrrr}
\hline
Node&\multicolumn{2}{c}{Choir}&&\multicolumn{2}{c}{Nave}&&\multicolumn{2}{c}{Nave}\\
&&&&\multicolumn{2}{c}{\scriptsize{FN1-FN2, FN5-FN11}}&&\multicolumn{2}{c}{\scriptsize{FN3-FN4}}\\
\cmidrule{2-3}\cmidrule{5-6}\cmidrule{8-9}
& $\delta_x$&$\delta_y$&&$\delta_x$&$\delta_y$&&$\delta_x$&$\delta_y$\\
\hline
\multicolumn{9}{c}{Own weight}\\
\hline
A&-0.17&-0.32&&0.03&-0.51&&-0.06&-0.04\\
B&-0.17&-1.27&&0.04&-0.71&&-0.04&-0.86\\
\hline
\multicolumn{9}{c}{Own weight+wind}\\
\hline
A&-16.60&-0.06&&-7.35&-0.02&&-7.56&0.07\\
B&-16.43&-5.13&&-7.12&-0.20&&-7.34&-0.41\\
\hline
\end{tabular}
\end{center}
\end{small}
\label{tab:7}
\end{table}

\subsubsection{Stresses}
\label{sec:stresses}
The load condition that gives the highest values of the stress in the wooden beams of the \cha\ is that of own weight plus wind in the case of the original state. In Tabs. \ref{tab:8}, \ref{tab:9} and \ref{tab:10}, the worst combination of internal actions, i.e. the one causing the highest value of the stress, is given for each different type of beam section, cf. Tab. \ref{tab:1}. $N$ is the axial force, positive when tension, $M_1$ and $M_2$ are the bending moments, $T$ the shear force, $\sigma_{max}$ the highest absolute value of the normal stress in the section, $\tau_{max}$ that of the shear stress and $\sigma_{eq}$ an upper bound of the equivalent stress of Von Mises calculated as:
\begin{equation*}
\sigma_{eq}=\sqrt{\sigma_{max}^2+3\tau_{max}^2}.
\end{equation*} 
Actually, because wood is not isotropic, an equivalent value of the stress should be calculated by a more appropriate criterion, e.g. the Hill's or the Hoffmann's one, \cite{vannucci}. Nevertheless, the values of the mechanical parameters of the ancient Notre-Dame's \cha\ are not known, so the use of such criteria is problematic. However, the stress level in the beams is always far below the admissible value $\sigma_{adm}$  for the failure of the oak wood normally accepted in the literature, say $\sigma_{adm}\sim58$ MPa for the axial force and $\sigma_{adm}\sim105$ MPa for bending, \cite{cirad}.
So, the possible inaccuracies dues to the inescapable incertitudes in the beams' dimensions and the approximation of $\sigma_{eq}$ do not affect substantially the check of the stress level in the \cha: without any doubt, the structure is rather little solicited. The most stressed pieces are the {\it jambes}: under the action of the wind, those on the leeward side are engaged to transmit the thrust to the lower part of the guttering wall, so they are particularly stressed, namely compressed and bent  by the {\it aisseliers}. This fact  attests of the real functioning and effectiveness of the system put in place by the carpenters  for equilibrating the wind action. 
 The {\it entraits} are less stressed: the system of the intermediary supports put in place by the builders is effective in reducing its bending. Also for the \cha\ FN3-FN4 of the nave, where  the link between the {\it suspente} and the {\it entrait} is broken, the stress remains very low, though it increases of  $\sim25\%$ with respect to that of the frames FN1-FN2, FN5-FN11. In the choir, the {\it entrait} is more stressed, $\sim8.5$ MPa, as a result of the different static scheme of the main frame, though it has a larger section than those of the nave's frames. This best static response of the nave's \cha\ is a consequence of the different static scheme and attests of its  better mechanic design.

\begin{table}[htp]
\caption{Internal actions and stresses in the choir's \cha, original state.}
\begin{small}
\begin{center}
\begin{tabular}{lrrrrrrr}
\hline
	&$N$   &       	$M_1$ &	$M_2$ 	&$T$     &    $\sigma_{max}$&	$\tau_{max}$ &$\sigma_{eq}$ \\
	&[N]&[N m] &[N m] & [N]& [MPa]& [MPa]& [MPa]\\
	\hline
	\multicolumn{8}{c}{Own weight}\\
	\hline
{\it Entrait}	&37422	&4590.06	&0	&1915&	1.106&	0.027&	1.107\\
1st {\it faux entrait}&	-2453&	2111.70&	0&	4298	&1.407&	0.184&	1.442\\
2nd {\it faux entrait}&	-12835	&640.49	&0	&187&	0.982&	0.008&	0.982\\
3rd {\it faux entrait}&	-2958	&1231.29&	0&	772	&1.017&	0.034&	1.018\\
4th {\it faux entrait}&	-1774	&177.97	&0&	0&	0.259&	0.000&	0.259\\
{\it Arbalétriers}&	-8103&	1666.67&	0&	104&	1.776&	0.005&	1.776\\
{\it Faux arbalétriers}&	-24633&	629.24&	0&	1346	&0.984&	0.042&	0.987\\
{\it Poteaux}&	14368&	2345.56	&0	&9240&	3.796&	0.486&	3.888\\
{\it Poteau central haut}&	2278	&0.24&	0&	0&	0.117&	0.000&	0.117\\
{\it Jambe gauche, jambettes}	&-5045	&1570.09	&0	&2416&	1.111&	0.088	&1.122\\
{\it Aisseliers j. gauche} and {\it liernes}	&-6988&	91.72	&0&	0&	0.399&	0.000	&0.399\\
{\it Jambe droite}&	-4934	&837.54	&0	&2323&	0.551	&0.061	&0.561\\
{\it Sablières}	&0&	1805.65	&3537.55&	2202	&7.109&	0.124&	7.112\\
{\it Liernes}	&0	&1451.47&	1.46&	3159&	0.921	&0.135	&0.950\\
\hline
\multicolumn{8}{c}{Own weight + wind}\\
\hline
{\it Entrait}	&76883	&47712.27&	0&	20523	&8.522&	0.293	&8.537\\
1st{\it  faux entrait}	&7383&	4611.30	&0	&4528&	3.130&	0.194&	3.148\\
2nd {\it faux entrait}&	11902	&3694.00	&0	&10418&	3.788&	0.471&	3.875\\
3rd {\it faux entrait}&	-6224&	5359.04&	0&	4839	&4.233&	0.210&	4.248\\
4th {\it faux entrait}	&-1991	&177.97	&0&	0	&0.267	&0.000&	0.267\\
{\it Arbalétriers}&	-2712	&1345.38&	0	&777&	1.322&	0.034&	1.323\\
{\it Faux arbalétriers}&	-29961&	1578.62&	0&	873&	1.800&	0.028&	1.801\\
{\it Poteaux}&	-9436&	4148.06&	0	&2499	&6.153&	0.132&	6.157\\
{\it Poteau central haut}	&-2113&	266.43&	0&	148&	0.690&	0.011&	0.691\\
{\it Jambe gauche, jambettes}	&83028&	24063.63&	0&	27403&	17.168&	0.993&	17.254\\
{\it Aisseliers j. gauche} and {\it liernes}	&105058	&91.72&	0	&0	&4.290&	0.000	&4.290\\
{\it Jambe droite}	&-12232&	131.55&	0	&363&	0.287&	0.010&	0.288\\
{\it Sablières}	&0&	2943.39&	6302.16&	3589	&12.224&	0.202&	12.229\\
{\it Liernes}&	0	&620.64	&4434.61	&0&	6.224&	0.000&	6.224\\
\hline
\end{tabular}
\end{center}
\end{small}
\label{tab:8}
\end{table}

\begin{table}[htp]
\caption{Internal actions and stresses in the nave's \cha, FN1-FN2, FN5-FN11, original state.}
\begin{small}
\begin{center}
\begin{tabular}{lrrrrrrr}
\hline
	&$N$   &       	$M_1$ &	$M_2$ 	&$T$     &    $\sigma_{max}$&	$\tau_{max}$ &$\sigma_{eq}$ \\
	&[N]&[N m] &[N m] & [N]& [MPa]& [MPa]& [MPa]\\
	\hline
	\multicolumn{8}{c}{Own weight}\\
	\hline
{\it Entrait}	&29222	&2399.38	&0	&12138	&1.046&	0.241	&1.126\\
{\it Faux entraits}&	-5704	&1107.67	&0	&4323&	0.819&	0.159	&0.864\\
{\it Arbalétriers}	&-20504	&937.66&	0	&2374	&1.043&	0.087	&1.054\\
{\it Faux arbalétriers}&	-18252&	441.47&	0&	2169&	0.997&	0.101&	1.012\\
{\it Poinçon}	&17586&	404.86&	0	&3803&	0.707&	0.131	&0.742\\
{\it Suspente} &	2912	&126.84&	0	&290&	0.321&	0.015&	0.322\\
{\it Poteaux}	&6788&	259.42&	0	&896	&0.429&	0.040&	0.434\\
{\it Jambettes}	&544	&371.19&	0&	1775	&0.603&	0.111&	0.633\\
{\it Liernes}	&0	&1178.40	&90.85&	3693&	1.589&	0.205&	1.629\\
{\it Aisseliers} and {\it blochets}	&-8124&	50.04&	0	&0&	0.450&	0.000&	0.450\\
{\it Jambes de force}	&-31443	&1777.49	&0	&14055&	3.418&	0.703&	3.628\\
{\it Chevrons secondaires}&	-6009	&1096.50&	0.02	&248&	0.819&	0.009&	0.819\\
{\it Sablières}	&0&	0&	728.81&	0	&0.630&	0.000&	0.630\\
\hline
\multicolumn{8}{c}{Own weight + wind}\\
\hline
{\it Entrait}&	41322	&12805.76	&0	&12138	&4.062&	0.241&	4.083\\
{\it Faux entraits}&	9001	&5752.51&	0&	4323	&3.745	&0.159&	3.756\\
{\it Arbalétriers}	&-37182&	4388.64	&0	&2374	&3.442	&0.087&	3.446\\
{\it Faux arbalétriers}	&-48625	&2562.83&	0&	2169&	4.011	&0.101&	4.015\\
{\it Poinçon}	&19250&	4030.64&	0	&3803	&3.450	&0.131	&3.457\\
{\it Suspente }&	6280	&812.21&	0	&290	&1.628&	0.015&	1.628\\
{\it Poteaux}	&-10080	&934.12&	0&	896&	1.121&	0.040&	1.123\\
{\it Jambettes}&	1170	&866.57&	0	&1775	&1.403	&0.111&	1.416\\
{\it Liernes}	&0&	237.72&	3198.68&	0	&5.032&	0.000&	5.032\\
{\it Aisseliers} and {\it blochets}	&-44331	&50.04	&0&	0&	2.059&	0.000&	2.059\\
{\it Jambes de force}&	-74798	&10073.06&	0	&14055&	15.924	&0.703&	15.970\\
{\it Chevrons secondaires}	&-5241&	2420.75&	30.91&	248	&1.638&	0.009&	1.639\\
{\it Sablières}&	0&	0&	1613.61&	0&	1.396&	0.000&	1.396\\
\hline
\end{tabular}
\end{center}
\end{small}
\label{tab:9}
\end{table}

\begin{table}[htp]
\caption{Internal actions and stresses in the nave's \cha, FN3-FN4, original state.}
\begin{small}
\begin{center}
\begin{tabular}{lrrrrrrr}
\hline
	&$N$   &       	$M_1$ &	$M_2$ 	&$T$     &    $\sigma_{max}$&	$\tau_{max}$ &$\sigma_{eq}$ \\
	&[N]&[N m] &[N m] & [N]& [MPa]& [MPa]& [MPa]\\
	\hline
	\multicolumn{8}{c}{Own weight}\\
	\hline
{\it Entrait}&	33433	&2574.57	&0	&365	&1.150&	0.007&	1.150\\
{\it Faux entraits}	&-8584&	1632.08	&0&	12362&	1.210&	0.454&	1.444\\
{\it Arbalétriers}&	-24786	&1163.32&	0&	1423&	1.278&	0.052	&1.282\\
{\it Faux arbalétriers}&	-21713&	428.42	&0	&370	&1.091&	0.017	&1.091\\
{\it Poinçon}	&18204&	238.63&	0&	295&	0.597&	0.010&	0.597\\
{\it Suspente }&	1759	&18.46&	0	&33	&0.093	&0.002&	0.093\\
{\it Poteaux}	&14825&	319.11	&0&	182	&0.718&	0.008&	0.718\\
{\it Jambettes}	&-400&	1068.34&	0	&3586&	1.686&	0.224&	1.730\\
{\it Liernes}	&0&	1239.67&	230.44&	5583	&1.872	&0.310&	1.947\\
{\it Aisseliers} and {\it blochets}	&-10699	&50.04&	0	&0	&0.564&	0.000&	0.564\\
{\it Jambes de force}	&-23169&	2367.41&	0	&3321&	3.929&	0.166&	3.939\\
{\it Chevrons secondaires}	&-3642	&1052.43&	0&	1374	&0.734&	0.051	&0.739\\
{\it Sablières}	&0&	349.36&	1462.54&	998&	1.479&	0.049&	1.482\\
\hline
\multicolumn{8}{c}{Own weight + wind}\\
\hline
{\it Entrait	}&47714	&15695.17	&0&	24293	&4.940&	0.483&	5.010\\
{\it Faux entraits}&	2141	&5977.86&	0&	5959	&3.715	&0.219	&3.735\\
{\it Arbalétriers	}&-35662	&4063.94&	0&	2126	&3.218&	0.078	&3.221\\
{\it Faux arbalétriers	}&-51853	&2321.21&	0&	2261&	3.875&	0.105&	3.879\\
{\it Poinçon	}&21351&	4329.90	&0	&4227	&3.721&	0.146&	3.730\\
{\it Suspente }	&-9944&	106.27&	0	&188&	0.530&	0.010&	0.530\\
{\it Poteaux	}&15951&	791.40&	0	&759&	1.167&	0.033&	1.169\\
{\it Jambettes}&	749&	732.49&	0&	1507	&1.176&	0.094&	1.187\\
{\it Liernes}	&0	&248.87&	3478.70&	0	&5.461&	0.000	&5.461\\
{\it Aisseliers }and {\it blochets}&	-50448&	50.04&	0	&0	&2.331&	0.000&	2.331\\
{\it Jambes de force}	&-66209&	11474.49&	0&	16008&	17.506&	0.800&	17.561\\
{\it Chevrons secondaires}&	-6098&	2460.42&	0	&300	&1.657&	0.011&	1.657\\
{\it Sablières}	&0&	0&	2709.11&	0	&2.344&	0.000&	2.344\\
\hline
\end{tabular}
\end{center}
\end{small}
\label{tab:10}
\end{table}

\subsection{About the transmission by friction of the horizontal forces}
\label{sec:horizforce}
As already mentioned, the simulations presented in the previous Sections are done in the assumption that the horizontal forces cannot be taken up by friction. In order to control this hypothesis and to evaluate the consequences of a link by friction between the \cha\ and the walls, we change the boundary conditions to the FE model of the \cha, taking for each one of the nodes S1 to S10 a fixed support, simulating a perfect contact, i.e. a support  with sufficient friction  to stop any sliding. This is  the condition normally assumed and implicitly understood when the horizontal thrust of the  \cha\ on the top of the guttering walls is supposed to exist. 

We bound the analysis to the original state of the \com\ and to the case of the structure loaded by its own weight. For such a loading case, we consider the total forces applied by a structural unit to the top of the guttering walls when all the supporting nodes are bilateral fixed supports, i.e., the displacements in the vertical and horizontal directions are completely blocked. The results of the simulations done in this way are shown in Tab. \ref{tab:11}, where $H$ is the total horizontal thrust of the \cha\ and $V$ the total vertical reaction, for each structural unit, applied to the top of the guttering walls. 

\begin{table}[htp]
\caption{Total forces, for each structural unit of \cha, on the top of the guttering walls in the assumption of fixed supports, original state (NW: North wall; SW: South wall).}
\begin{small}
\begin{center}
\begin{tabular}{lrrrrr}
\hline
&\multicolumn{2}{c}{Choir}&&\multicolumn{2}{c}{Nave}\\
\cmidrule{2-3}\cmidrule{5-6}
&SW&NW&&SW&NW\\
\hline
$H$ \scriptsize{[N]}&46048&45356&&40790&43389\\
$V$ \scriptsize{[N]}&68030&68354&&63612&63056\\
$V_w$\scriptsize{[N]}&160000&160000&&136100&136100\\
$V_t$\scriptsize{[N]}&228030&228354&&199712&199156\\
$M$ \scriptsize{[N m]}&124330&122461&&110133&117150\\
$e$ \scriptsize{[m]}&0.545&0.536&&0.551&0.588\\
\hline
\end{tabular}
\end{center}
\end{small}
\label{tab:11}
\end{table}

We check  the global equilibrium of the guttering wall 2.70 m below its top, i.e. at the same level of the corbels supporting the timber {\it consoles} of the \cha. This was the level of the top of the clerestory walls before the modifications started around 1220 and it is, to a good approximation, the level at which the vault touches the clerestory walls. In other words, we can consider the free standing height of the guttering walls to be of the order of 2.7 m.  If we consider a thickness of the guttering walls of 60 cm and a density of the limestone of 2400 kg/m$^3$, the weight $V_w$ of this part of the guttering wall is of $\sim160000$ N for the choir (length of 4.1 m) and of $\sim136100$ N for the nave (length of 3.5 m). The total vertical load $V_t$ at the level $-2.70$ m with respect to the top of the wall can hence be calculated, as well as the overturning moment $M$ of the horizontal thrust, and finally the eccentricity $e$ of  $V_t$ with respect to the centroid of the wall's section, cf. Tab. \ref{tab:11}. The values of the eccentricity $e$ so calculated, greater than 50 cm for all the cases, are extremely high and should cause the overturning of the wall at the level of the ancient top of the clerestory wall, the one before the changes of 1220. Also if the wall had a greater thickness, the eccentricity should be too large to ensure a safe equilibrium of the system \cha-guttering walls. 

In addition, the physical possibility for the system to develop effective friction forces should be investigated. To this end, let us consider a friction coefficient $\nu=0.7$ for the contact between wood and stone, value usually admitted in such a  case for  a dry contact. In Tab. \ref{tab:12} the values of the horizontal, $R_x$, and vertical, $R_y$, contact forces on the top of the guttering walls, still calculated in the  assumption of fixed bilateral supports, are shown for each frame, cf. Fig. \ref{fig:17}. It is apparent that the ratio $R_x/R_y$ exceeds $\nu$ in several cases, while it is close to it in other cases. In addition, the contact between the wood of the \cha\ and the stone of the guttering walls is probably far from being perfect: infiltrations of dust and  rain water cannot be excluded, especially if one considers that the contact is actually unilateral and that, as shown in the numerical simulations presented in Sect. \ref{sec:reactions}, some nodes of the \cha\ can slightly lift up under the action of the loads, cf. Figs. \ref{fig:23} and \ref{fig:24}. 

Finally, the transmission, by friction, of horizontal forces between the \cha\ and the top of the guttering walls should be  not only dangerous for the  safety of the structure, but also rather uncertain or even impossible, physically speaking. The masterbuilders  of the XIIIth century needed hence  to transfer the horizontal forces of the \cha\ to the stone structure below in another way. For what concerns the horizontal thrusts produced by the own weight of the structure at the base of the {\it chevrons}, they ideated a horizontally self equilibrated system, composed by the {\it chevrons maîtres}, equipped with a tie, the {\it entrait}, the {\it sablières} and {\it blochets} for the support of the secondary {\it chevrons} and the bracing system, composed by longitudinal {\it liernes} and {\it aisseliers}, to transfer the maximum of the load from the {\it fermettes} to the {\it fermes principales}. For transferring the wind thrust, they used the system of {\it consoles}, that operate the transfer of the force from the \cha\ to the wall by contact, on the leeward side, to a lower level with respect to the top of the guttering wall, so as to preserve the upper part of them from the overturning that a horizontal force applied on their top could produce. This same type of contact mechanism was able to safely equilibrate the flying-buttresses thrust during the construction phases, before the erection of the high vault.

\begin{table}[htp]
\caption{Reaction forces on the top of the guttering walls in the assumption of fixed supports, original state; $R_x$: horizontal reaction, $R_y$: vertical reaction.}
\begin{small}
\begin{center}
\begin{tabular}{lrrrrrrrrr}
\hline
Frame&Wall&\multicolumn{3}{c}{Own weight}&&\multicolumn{3}{c}{Own weight + wind}\\
\cmidrule{3-5}\cmidrule{7-9}
&&$R_x$&$R_y$&$R_x/R_y$&&$R_x$&$R_y$&$R_x/R_y$\\
&&[N]&[N]&&&[N]&[N]\\
\hline
\multicolumn{9}{c}{Choir}\\
\hline
\multirow{2}*{F1}&North&22058&	26586&	0,83&&	30368&	31337&	0,97\\	
			 &South&-21342&	26604&	0,80&&	-12317&	22293&	0,55\\
\multirow{2}*{F2 and F3}&	North&5983&	10378&	0,58&&	8933	&	10650&	0,84\\
				     &South&-5994&	10339&	0,58&&	-3232&	11044&	0,29	\\
\multirow{2}*{F4 and F5}&	North&6012&	10506&	0,57&&	9254&	11570&	0,80\\
				     &South&-6013&	10374&	0,58&&	-2961&	10185&	0,29\\
\hline
\multicolumn{9}{c}{Nave}\\
\hline
\multirow{2}*{F1}&North&	14523	&21722&	0.67&&	26051	&28120	&0.93\\
&South	&-15064	&22122	&0.68&&	-3663&	16145&	0.23\\
\multirow{2}*{F2 and F3}&	North	&8119&	10281&	0.79	&&9216&	13483&	0.68\\
&South&	-6490	&10338&	0.63	&&-3995&	7997	&0.50\\
\multirow{2}*{F4 and F5}&	North	&6314	&10386&	0.61&&	9137&	13796&	0.66\\
&South&	-6373	&10407	&0.61&&	-3808	&7788	&0.49\\
\hline
\end{tabular}
\end{center}
\end{small}
\label{tab:12}
\end{table}

\subsection{Modal analysis of the \textit{charpente}}
\label{sec:dynamic}

The structural improvements done on the \cha\ of the nave with respect to that of the choir can be indirectly, but rather explicitly, assessed by comparison of the modal analysis of the two \cha$s$. In fact, the weight of the structural units of the two \cha$s$ is practically the same, cf. Tab. \ref{tab:6}. Hence, a comparison of the fundamental frequencies of the two structures gives a rather good appraisal of their respective stiffnesses: the higher the frequency, the higher the stiffness.

A comparison of the frequencies of the first five fundamental modes is presented in Tab. \ref{tab:13}. The evaluation has been done with  the same assumptions used for the check of the friction mechanism, in the previous Section: all the nodes from S1 to S10 are fixed bilateral supports. Actually, to perform a  nonlinear dynamical simulation with unilateral contacts is out of the scope of this study while a simulation done with fixed supports remains sufficient to compare the two structures to have an assessment of their respective stiffnesses.

What is apparent from the results shown in Tab. \ref{tab:13} is that the nave's structural unit has a stiffness sensibly greater than that of the choir's one. Comparing the fundamental frequencies, mode 1, we can see that  the frequency of the nave's structural unit is $\sim2.5$ times that of the choir's one. This is a clear, tangible sign of the better structural conception of the nave's \cha\ with respect to that of the choir. The frequencies of the structural units of FN3 and FN4 are slightly lower than those of the other parts of the nave's \cha\, for the differences between these structural units of the nave's \com, already described in Sect. \ref{sec:mechmod}, differences that have the global effect of decreasing the stiffness of this part of the nave's \cha.

\begin{table}[htp]
\caption{Vibration frequencies of the choir's and nave's \cha, [Hz].}
\begin{small}
\begin{center}
\begin{tabular}{lrrrrrrrr}
\hline
Mode&\multicolumn{2}{c}{Choir}&&\multicolumn{2}{c}{Nave}&&\multicolumn{2}{c}{Nave}\\
&&&&\multicolumn{2}{c}{\scriptsize{FN1-FN2, FN5-FN11}}&&\multicolumn{2}{c}{\scriptsize{FN3-FN4}}\\
\cmidrule{2-3}\cmidrule{5-6}\cmidrule{8-9}
&OS&FS&&OS&FS&&OS&FS\\
\hline
1&0.206&0.228&&0.521&0.569&&0.464&0.464\\
2&0.215&0.239&&0.607&0.666&&0.521&0.569\\
3&0.632&0.696&&1.142&1.209&&0.606&0.665\\
4&0.648&0.714&&1.255&1.260&&1.098&1.204\\
5&0.687&0.774&&1.747&1.350&&1.142&1.259\\
\hline
\end{tabular}
\end{center}
\end{small}
\label{tab:13}
\end{table}

\begin{figure}
\begin{center}
\includegraphics[width=\textwidth]{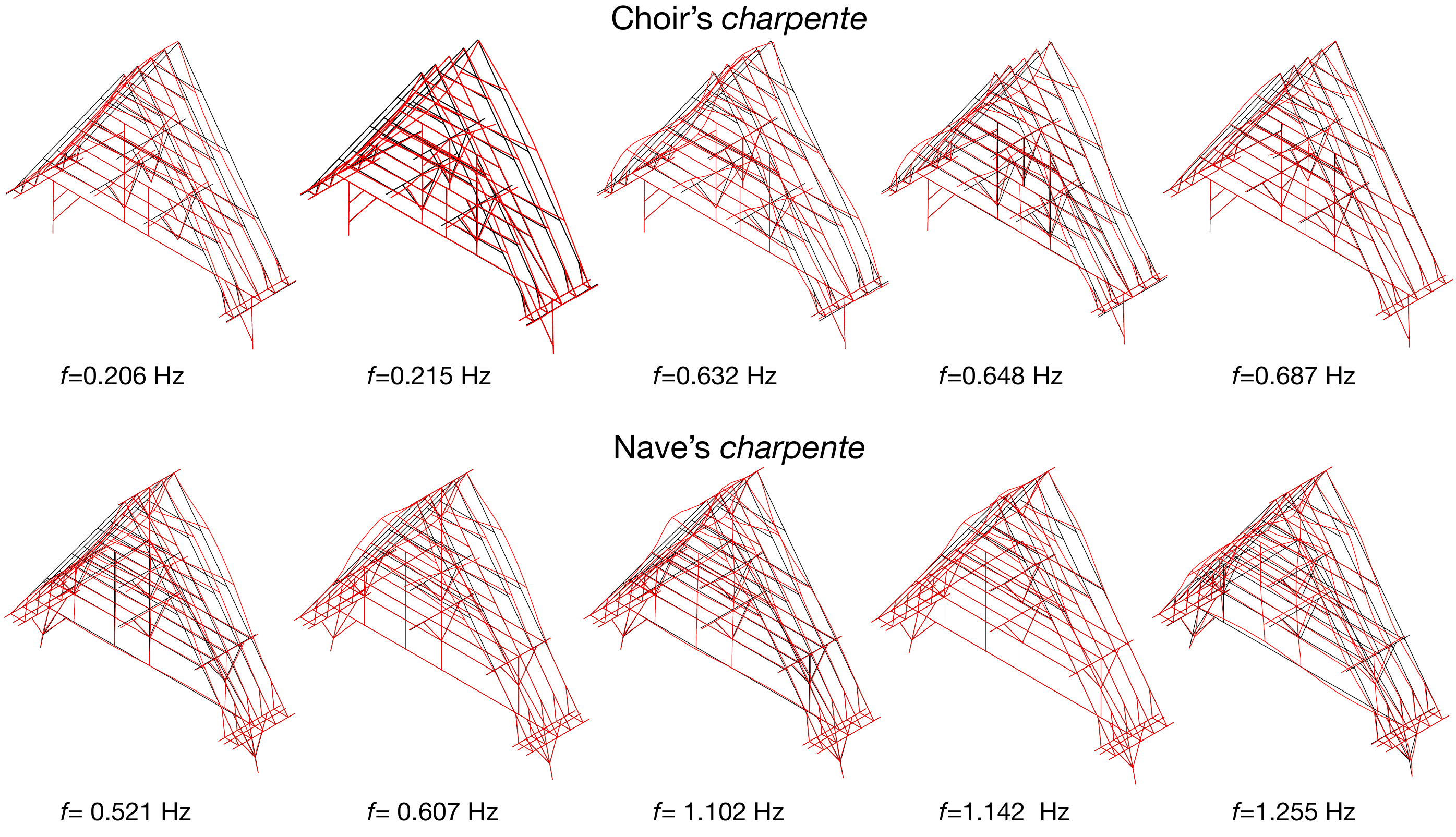}
\caption{First five normal modes and corresponding vibration frequencies, original state.}
\label{fig:25}
\end{center}
\end{figure}

\section{Final considerations and conclusion}
\label{sec:interpret}
The results presented in this paper confirm some points already known about the timber \com\ of the Gothic age, namely of those built on the scheme of the {\it chevrons formant fermes}, and suggest some different interpretations, according to the quantitative analysis based upon the numerical simulations conducted on a FE model of the Notre-Dame's ancient  \cha. The main observations that can be drawn are summarized hereafter.

On the whole, we have seen that the structural behavior of a \cha\ with {\it chevrons formant fermes} is rather complicate and that it cannot be analyzed  by a simple planar scheme: the behavior of such a type of structure is strongly three-dimensional. The fact that the masterbuilders of the Middle Ages could invent and use this structural solution attests their deep comprehension of the mechanical functioning of these structures. 

From the three-dimensional structural analysis of the \com, some interesting points emerge. First of all, the determinant role of the {\it fermes principales} that, thanks to the longitudinal bracing system of {\it liernes} and {\it aisseliers}, take on the largest part of the vertical loads, as the numerical simulations clearly show. The usual idea that the {\it chevrons formant fermes} were used to distribute the vertical load almost uniformly on the top of the clerestory walls does not correspond to reality and seems to be dictated by a simple two-dimensional analysis of the structure or also, perhaps, suggested by ideological positions, both of them far from the physical reality.

Then, the carpenters well understood the need for using an effective and safe system to transmit the horizontal forces on the \cha\ to the stone structure below. For what concerns the horizontal thrust produced by the own weight of the structure, the global three-dimensional functioning of the {\it chevrons formant fermes} corresponds to that of a structure that is globally self-equilibrated in the horizontal direction. The thrusts at the bases of the {\it chevrons} of the {\it fermettes} are transmitted to the main frame through the system of {\it sablières} and {\it blochets} and also by the longitudinal bracing system. The {\it entraits} of the main frames ensure then the equilibrium absorbing the thrusts on the opposite sides, in this only partially helped by the  {\it consoles}. 

These last constituted, in reality, the device used to transfer, by contact, the horizontal action of the wind to the lower part of the guttering wall on the leeward side. The Gothics realized in this way an effective system, transforming the wind thrust into an inclined force applied as low as possible, which preserves the guttering wall from overturning. The same system of {\it consoles} was effective during the construction of the high vault, to counterbalance the thrust of the flying-buttresses, while the common idea that the consoles where used to sustain the {\it fermes principales} and to help the {\it entraits} in bending seems to be illusory. This system of transfer of the horizontal wind force is with no doubt to be put in relation with the raise of the guttering walls occurred around 1220. Such a change allowed the masterbuilders to adopt a new scheme for the \cha, namely using the {\it entraits} and hence allowing the global three-dimensional functioning of the \cha\ described above. Whether or not the guttering walls were raised just for adopting a better scheme for the \cha\ as consequence of structural problems occurred to the \cha\ of the XIIth century cannot be affirmed with certitude, nor  excluded {\it a priori}. In all the cases, the common idea that the {\it chevrons} applied a horizontal thrust on the top of the guttering walls by friction is not corroborated by the mechanical analysis and all seems to indicate that for the Gothics this had not to happen: the system of {\it sablières-blochets} and the {\it consoles} are clear signs of this. 

Undoubtedly the solution of a \cha\ with {\it chevrons formant fermes} allowed the use of  trunks of moderate diameter, easier to be found in the France of the Middle Ages, so realizing relatively light structures, though this seems more properly dictated by economical reasons than by statics. In fact, the dimensions used for the timber beams are largely sufficient to withstand the loads and a heavy load on the top of the clerestory walls could help in stabilizing the whole construction, knowing that the stress level in the stone of a Gothic cathedral is normally far from being critical, \cite{heyman}.
Anyway, the change of the static scheme from the choir's \cha\ to the nave's one was certainly not dictated by economical reasons, because the mass of wood for unit length of longitudinal axis is greater for the nave. What clearly appears from a study of the two \cha$s$ is that the nave's one is  better designed and that this results in an increased stiffness of the wall structure. This is confirmed by a comparative modal analysis of the two \cha$s$: the structural unit of the nave has a fundamental frequency which is much higher than that of the choir, for about the same mass. 

Nowhere in the \com\ the stresses reached important values: all the structure was feebly solicited, thanks to its peculiar design. All seems indicate that the Gothics were guided by the need of increasing the stiffness of the structure, a notion with no doubts easier to be grasped  empirically than that of stress, especially thanks to the construction phases. A non secondary role, not investigated here, is perhaps that of the {\it voligeage}, probably determining in preventing from local elastic buckling phenomena in the compressed and bent slender {\it chevrons}.

The ancient \cha\ of Notre-Dame, destroyed by the fire of April 15th, 2019, will be reconstructed. At present, it is not yet clear how and with which technology. We can anyway learn much from the ancients. They were able to construct a structure that without major problems lasted almost intact more than eight centuries and perhaps it would have last much longer, in consideration of its good structural behavior and of the lower level of the stresses. The work and ideas of the Gothic carpenters can suggest us how to realize a hopefully so beautiful and long-lasting structure, which is uncommon for modern engineering. Care should be taken to the structural scheme to be used and mainly to the way the forces are transferred to the stone structure of the clerestory, a problem  brillantly solved by the Gothics.



\bibliographystyle{plainnat} 
\bibliography{bibliocharpente}   

\end{document}